\documentclass[useAMS]{mn2e}
\usepackage{graphicx}
\usepackage{amsmath, amssymb}

\begin{document}

\title[Tidal Disruptions and Black Hole Recoil]{Tidal Disruption Flares of Stars from Moderately Recoiled Black Holes}
\author[N. Stone and A. Loeb]{Nicholas Stone$^1$\thanks{nstone@cfa.harvard.edu} and Abraham Loeb$^1$\thanks{aloeb@cfa.harvard.edu}\\
$^1$Harvard-Smithsonian Center for Astrophysics, 60 Garden Street, Cambridge, MA 02138, USA}

\maketitle
\begin{abstract}
We analyze stellar tidal disruption events as a possible observational signature of gravitational wave induced recoil of supermassive black holes.  As a black hole wanders through its galaxy, it will tidally disrupt bound and unbound stars at rates potentially observable by upcoming optical transient surveys.  To quantify these rates, we explore a broad range of host galaxy and black hole kick parameters.  We find that emission from a transient accretion disk can produce $\sim 1$ event per year which LSST would identify as spatially offset, while super-Eddington tidal flares, if they exist, are likely to produce $\sim 10$ spatially offset events per year.  A majority of tidal disruption flares, and a large majority of flares with an observable spatial offset, are due to bound rather than unbound stars.  The total number of disruption events due to recoiled black holes could be almost $1\%$ of the total stellar tidal disruption rate.

\end{abstract}
\begin{keywords}
Tidal Disruption Flares -- Black Hole Recoil
\end{keywords}

\section{Introduction}
Recent advances in numerical general relativity quantified how the  coalescence of unequal black hole binaries leads to the anisotropic emission of gravitational radiation, which can carry enough linear momentum to deliver a substantial kick to the merged black hole (Pretorius 2005, Baker et al. 2006, Campanelli et al. 2006).  Because the inspiral and eventual coalescence of supermassive black holes (SMBHs) is a frequent consequence of galaxy mergers, it is expected that SMBH recoil will often accompany these events.  The magnitude of the recoil varies as a function of the initial mass ratio and the spin and orbital angular momentum vectors of the coalescing black holes, but is generally of order hundreds of kilometers per second (Schnittman \& Buonanno 2008, Lousto et al. 2010).  At the high end of the velocity distribution, the black hole can recoil with a velocity $\gtrsim 1000 \rm ~km~s^{-1}$ and escape the merged galaxy.  More commonly, it will oscillate for a time ranging from $10^6$ - $10^9 {\rm ~yrs}$ (Blecha \& Loeb 2008, Sijacki et al. 2010, Blecha et al. 2010) before settling down into the galaxy's center.

Observation of a recoiling SMBH would be of interest both as a probe of hierarchical galaxy evolution and as a test of the predictions of strong field general relativity.  Consequently, several papers have focused on observational signatures of a recoiling SMBH.  A black hole kicked at a substantial fraction of its host galaxy's escape velocity can create or expand a low-density stellar core (Gualandris \& Merritt, 2008).  The small cloud of stars gravitationally bound to the recoiling black hole would appear like a globular cluster, but with much higher velocity dispersion (O'Leary \& Loeb 2009, Merritt et al. 2009, O'Leary \& Loeb 2011). Gas accretion onto the black hole, manifested as a spatially or kinematically offset quasar (Madau \& Quataert 2004, Loeb 2007), is potentially a very clear signature, but the gas reservoir bound to the black hole will be depleted within $10^7$ years after ejection from the center of its galaxy (Blecha \& Loeb 2008, Blecha et al. 2010).  Several candidates have already been detected with this last method (Komossa et al. 2008, Shields et al. 2009, Civano et al. 2010), although the evidence for black hole recoil is not conclusive (Bogdanovic et al. 2009).  A potentially longer-lived source of accretion power is the tidally disrupted matter from stars passing too close to the recoiling black hole, which could be visible as an off-center tidal disruption flare.  Because tidal disruption flare lightcurves have, in principle, several unique identifiers (Strubbe \& Quataert 2009, Guillochon et al. 2009, Kasen \& Ramirez-Ruiz 2010, Strubbe \& Quataert 2010), an off-center or intergalactic tidal disruption flare would be a very strong indication of a recoiling black hole.  This scenario was first investigated in a paper by Komossa \& Merritt (2008, hereafter KM08).  However, to evaluate the utility of tidal disruption signatures for recoiling black holes it is first necessary to obtain estimates for the tidal disruption rates produced by these systems.  KM08 calculate these rates for relatively large kick velocities, while in this paper we generalize the calculation to cover realistic, and often moderate, kick velocity distributions.  The aim of these calculations is to approximate the frequency of off-center and spectrally shifted tidal disruption flares, to determine if tidal disruption events (TDEs) are useful probes of physically motivated kick velocity distributions.  These rate calculations are also relevant to SMBHs ejected from galaxy centers in triple-SMBH interactions (Hoffman \& Loeb 2007), although we do not attempt to evaluate the more complicated distributions of parameters for such events.

Very large increases in tidal disruption rates have been discussed previously in the context of galaxy mergers, both as a prompt electromagnetic signal immediately following SMBH coalescence (Stone \& Loeb 2010), and due to resonances or chaotic stellar orbits at the end of the dynamical friction phase of an SMBH binary (Ivanov et al. 2005, Chen et al. 2009, Wegg \& Bode 2010, Chen et al. 2011).  Neither of these mechanisms is capable of producing a tidal disruption flare with an observable spatial offset, as both can only occur in galactic nuclei.  Tidal flares produced by wider SMBH binaries could still be a source of confusion in interpreting spatially offset TDEs, however.  Binary-produced flares would themselves be of interest, but would not be directly useful for testing predictions of gravitational wave recoil.  Distinguishing between the two possibilities will be simple in some cases: for instance, a TDE in intergalactic space, independent of an observable stellar population, would likely be due to a recoiling black hole; whereas a TDE with periodic interruptions in its lightcurve (Liu et al. 2009) would be due to a hard SMBH binary.  However, in many cases it will be not be trivial to disentangle the two causes, and detailed observation and modeling of the host galaxy could be necessary to determine if it is likely to harbor a binary SMBH at the observed spatial or kinematic offset.  A final source of confusion could be disruption of stars by IMBHs left over in a galaxy's halo from earlier stages of hierarchical growth (O'Leary \& Loeb 2011).  However, these TDEs would probably be distinguishable due to their low black hole masses, and in any event would be a complementary example of black hole recoil.

Although the current sample of observed TDEs is small, with roughly a dozen strong candidates (Gezari et al. 2009), current and upcoming time-domain optical transient surveys such as Pan-STARRS\footnote{http://pan-starrs.ifa.hawaii.edu/public/}, PTF\footnote{http://www.astro.caltech.edu/ptf/}, and LSST\footnote{http://www.lsst.org/lsst} are expected to increase that sample by $1-3$ orders of magnitude (Strubbe \& Quataert 2009), making it profitable to study sub-populations of TDEs.  If sufficiently large, the subset of disruption flares associated with recoiling black holes could be used to constrain the LISA\footnote{http://lisa.nasa.gov/} event rate or the distribution of kick velocities associated with astrophysical SMBH mergers; at the very least, detection of this subset of TDEs could qualitatively confirm recent numerical relativity predictions.

The outline of this paper is as follows.  In \S 2 we develop the model used to estimate the TDE rate over a wide range of kick velocities and galaxy parameters, and in \S 3 we explain in more detail the distributions of disruption properties we integrate over.  In \S 4 we discuss the results of our modeling, and in \S 5 we offer our conclusions on the viability of tidal disruption flares as a technique for identifying recoiled SMBHs.

\section{Model}
The three primary quantities to be calculated are the trajectory of the kicked black hole through its host galaxy, the rate at which it disrupts unbound stars that it encounters, and the rate at which it depletes its cloud of gravitationally bound stars via tidal disruption.  The location (and possible kinematic offset) of a disruption event relative to its host galaxy will determine whether the flare can be distinguished from a TDE due to a stationary black hole.  \S 2.1 examines expected ranges of black hole kicks, \S 2.2 discusses the relevant tidal disruption physics, \S 2.3 lays out the galaxy parametrization used in this paper, and \S 2.4 and \S 2.5 describe the techniques used to estimate TDE rates for unbound and bound stars, respectively.  Finally, in \S 2.6 we discuss observational constraints.  Our general strategy is as follows: Strubbe \& Quataert 2009 (hereafter SQ09) calculate the number of TDEs a generic survey would be expected to detect, binned by black hole mass (see SQ09 Figures 9, 13).  In this calculation they assume a time-averaged tidal disruption rate $\dot{N}=10^{-5}\rm ~yr^{-1}~galaxy^{-1}$.  Our model averages over a kick velocity distribution and a range of galaxy parameters (detailed in \S 2.3) to calculate a $\dot{N}$ dependent on black hole mass, and then modifies the SQ09 survey calculations accordingly.

\subsection{Black Hole Kicks}
The gravitational recoil velocity of a post-merger SMBH depends only on the mass ratio, spin amplitudes, and spin orientations (relative to the orbital angular momentum plane) of the two progenitor SMBHs.  The merger of Schwarzschild black holes represents a simple case, with a maximum kick velocity of $\sim$175 ${\rm km~s^{-1}}$ occurring at a mass ratio of 0.36 (Gonzalez et al. 2007).  As the dimensionless spin parameters of the two SMBHs $a_1$, $a_2$ increase, so does the maximum kick velocity $v_{\rm k}$, although it has complicated functional dependences on the relative inclination of the pre-merger spin axes and the orbital angular momentum axis.  An exact calculation requires the full framework of numerical relativity, but analytic fitting formulae can be calibrated to numerical relativity results, recently yielding quite high accuracy (Lousto et al. 2010).  Although certain combinations of SMBH binary initial parameters can produce kicks up to $\sim 4000~ {\rm km~s^{-1}}$ (Campanelli et al. 2006), observation constrains how frequent such high-velocity kicks can be, due to the fact that galaxies with bulges all seem to possess central SMBHs (Ferrarese \& Ford 2005, Blecha et al. 2010).  This constraint is weakened, however, by the hierarchical nature of structure formation (Schnittman 2007).  

Higher velocity kicks become more important if one assumes high values of pre-merger spin amplitudes: Lousto et al. (2009) found 23\% of their Monte Carlo sample to exceed $1000~{\rm km~s^{-1}}$ by assuming all pre-merger spin amplitudes $a_1=a_2=0.97$ (with isotropic distribution of spin angles), and mass ratios $q$ between $1$ and $1/16$.  Spins of this magnitude are likely realistic for a nontrivial fraction of astrophysical black holes: the Fe $\rm K\alpha$ line has implied a near-maximally spinning SMBH candidate (Brenneman \& Reynolds 2006).  However, there is reason to believe that the spins of merging black holes may align during the inspiral phase (Bogdanovic et al. 2007) in the case of gas-rich mergers.  This has the potential to strongly suppress high-velocity kicks relative to an isotropic spin distribution (but see King et al. 2005 and Lodato \& Pringle 2006 for a description of how counter-alignment, which does not suppress $v_{\rm k}$ as strongly, can also occur).  Dotti et al. (2009) find a dramatic reduction, with median kick velocities below $70~{\rm km~s^{-1}}$, although their scenario investigates the specific case where $q$ is very near 1, while we are concerned with a wider range of mass ratios.  Their results also indicate that cooler gas accretion (adiabatic index $\gamma=5/3$) more effectively aligns spins and suppresses kicks than a warmer ($\gamma=7/5$) accretion flow.  More recently, it was discovered that even in gas-free mergers, relativistic spin precession is capable of aligning or anti-aligning progenitor SMBH spin vectors, and reducing median kick velocities (Kesden et al. 2010).  Consideration of all these factors highlights the need to generalize the work of KM08 to a more realistic distribution of kick velocities, as recoil velocities in excess of galactic escape speed are likely to be uncommon even if SMBH spins are near-maximal and unaligned.  

Because the real distribution of kick velocities depends on the SMBH spin distribution and the uncertain physics of the last stages of SMBH merger, we calculate black hole trajectories given kicks of $100, 200, 300, 400, 500, 600, 700, 800$, and $900$ $\rm~km~s^{-1}$, and then interpolate quantities of interest such as wandering lifetime and time-averaged tidal disruption rates.  We then fit these functions of $v_{\rm k}$ to plausible black hole kick distributions.  In particular, we consider the three $v_{\rm k}$ distributions in Figure 2 of Lousto et al. 2010, which assume isotropic pre-merger spins, spins aligned to within $30$ degrees, and spins aligned to within $10$ degrees (dry mergers, hot wet mergers, and cold wet mergers, respectively).  These scenarios assume a $q$ uniformly sampled between 0 and 1.  The dry merger scenario assumes spin magnitudes randomly sampled between 0 and 0.9, while the wet mergers assume spin magnitudes between 0.3 and 0.9.  In reality, the distribution of spin amplitudes also depends on factors not considered; chiefly, relativistic spin precession (Kesden et al. 2010) and chaotic versus standard accretion (Berti \& Volonteri 2008).  Likewise, modeling SMBH merger histories and spin evolution using the Press-Schechter formalism indicates a top-heavy spin distribution and a bottom heavy $q$ distribution (Volonteri et al 2004).  We leave more complicated $a_1$, $a_2$, and $q$ distributions to future research, although the kick distributions we consider should bracket a wide range of the available parameter space. 

\subsection{Tidal Disruption Physics}

Stars that pass within a radius
\begin{equation}
r_{\rm t}=r_*\left(\frac{\eta^2M_{\rm BH}}{m_*}\right)^{1/3}
\end{equation}
of a SMBH will be tidally disrupted.  In this equation $M_{\rm BH}$ is the black hole mass, $m_*$ and $r_*$ are the stellar mass and radius, and $\eta$ is a stellar structure constant which is of order unity for main sequence stars.  Detailed calculations find $\eta=0.844$ for $n=3$ polytropes, for example (Diener et al. 1995).  During the disruption event, roughly one half of the stellar mass is ejected from the system, while the other half remains bound (Rees 1988) but with the change in gravitational potential across the star producing a wide spread of specific orbital energies,
\begin{equation}
\Delta \epsilon \approx \frac{GM_{\rm BH}}{r_{\rm p}}\frac{r_*}{r_{\rm p}}.
\end{equation}
Here $r_{\rm p}$ is the pericenter distance of the star on its orbit around the black hole.  The bound stellar matter quickly expands to the point where hydrodynamic forces can be neglected and the gas follows roughly Keplerian trajectories, but upon return to pericenter the gas streams dissipate energy in shocks and form an accreting torus.  The characteristic mass return rate is 
\begin{equation}
\dot{M}_{\rm r}=\frac{1}{3}\frac{m_*}{t_{\rm r}}\left(\frac{t}{t_{\rm r}}\right)^{-5/3}
\end{equation}
(Phinney 1989), with the return time for the most tightly bound debris being
\begin{equation}
t_{\rm r}\sim\frac{2\pi}{6^{3/2}}\left(\frac{r_{\rm p}}{r_*}\right)^{3/2}\left(\frac{r_{\rm p}^3}{GM_{\rm BH}}\right)^{1/2}
\end{equation}
(SQ09).  This simple dynamical picture has been confirmed as largely accurate by numerical hydrodynamic simulations (Evans \& Kochanek 1989).  The radiative properties of the flare have been modeled by Loeb \& Ulmer (1997), and most recently by SQ09, whose model finds that mass infall rates can exceed the Eddington limit for days to months after disruption.  This super-Eddington phase may produce an outflow with supernova-like optical luminosities.  After super-Eddington infall ceases, emission is dominated by a thermal component from the accretion disk, which also photoionizes the unbound stellar debris, producing broad emission lines.  Luminosity from the disk fades with a decay time of order months to years, and is peaked in the UV and soft X-ray.  Upcoming transient surveys will dramatically expand the number of TDEs available for study, with the Pan-STARRS $3\pi$ survey expected to find $\sim10$ per year, rising to $\sim200$ if the SQ09 predictions about super-Eddington outflows are correct.  LSST is expected to find $\sim100$ per year, or up to $\sim6000$ if the hypothesized super-Eddington outflows can be seen.  In this paper we consider a black hole mass range from $10^6 M_{\sun}$ to $10^8 M_{\sun}$.   For $M_{\rm BH} \ga 10^8 M_{\sun}$ the Schwarzschild radius of a black hole exceeds the tidal disruption radius of main sequence stars, meaning that stars will be swallowed whole rather than disrupted by non-spinning SMBHs.  However, for a Kerr black hole $r_{\rm t}$ is angle-dependent, and for high values of spin, black holes as large as $\sim 7\times10^8 M_{\sun}$ can still tidally disrupt main sequence stars which approach from angles near the equatorial plane (Beloborodov et al. 1992).  For simplicity we consider SMBHs up to but not above $10^8 M_{\sun}$.  

In this paper we follow the prescriptions of SQ09 and Strubbe \& Quataert 2010 for disk and super-Eddington outflow luminosity.  Specifically, we model the super-Eddington outflow as a sphere with photosphere radius 
\begin{equation}
R_{\rm ph}\sim 10f_{\rm out}f_{\rm v}^{-1}\left(\frac{\dot{M}_{\rm r}}{\dot{M}_{\rm Edd}}\right)R_{p, 3R_{\rm S}}^{1/2}R_{\rm S} \label{RPhoto},
\end{equation}
photosphere temperature
\begin{align}
T_{\rm ph} \sim &2\times 10^5\left(\frac{f_{\rm v}}{f_{\rm out}}\right)^{1/3} \left(\frac{\dot{M}_{\rm r}}{\dot{M}_{\rm Edd}}\right)^{-5/12}\label{TPhoto} \\
& \times M_{6}^{-1/4} R_{p, 3R_{\rm S}}^{-7/24} ~{\rm K}, \notag
\end{align} 
and time of peak emission equal to the larger of $t_{\rm r}$ and the time when the outer edge of the photosphere becomes optically thin, 
\begin{equation}
t_{\rm edge} \sim 1~ f_{\rm out}^{3/8}f_{\rm v}^{-3/4}M_{6}^{5/8}R_{p, 3R_{\rm S}}^{9/8}m_*^{3/8}r_*^{-3/8}~ {\rm days}\label {tEdge}.
\end{equation}
Here the outflowing wind velocity is given by
\begin{equation}
v_{\rm w}=f_{\rm v}\left(\frac{GM_{\rm BH}}{r_{\rm p}}\right)^{1/2},
\end{equation}
with $f_{\rm v}$ a free parameter of fiducial value $1$, and the total mass flux in the outflow is
\begin{equation}
\dot{M}_{\rm out} = f_{\rm out}\dot{M}_{\rm r},
\end{equation}
with $f_{\rm out}$ a parameter taken to be $0.1$.  $\dot{M}_{\rm Edd}$ is the SMBH's Eddington-limited mass accretion rate assuming an accretion efficiency of $0.1$, $M_6=M_{\rm BH}/(10^6M_{\odot})$, $R_{\rm S}=2GM_{\rm BH}/c^2$, and $R_{p, 3R_{\rm S}}=r_{\rm p}/(3R_{\rm S})$.

Again following SQ09, we approximate the disk as a geometrically slim multicolor blackbody extending from the innermost stable circular orbit, $R_{\rm ISCO}$, to $2r_{\rm p}$.  Its temperature profile is given by
\begin{align}
\sigma &T_{\rm eff}^4(R) = \frac{3GM_{\rm BH}\dot{M}_{\rm r}f}{8\pi R^3}\times\label{diskEmit} \\
&\left(\frac{1}{2}+\sqrt{\frac{1}{4}+\frac{3}{2}f\left(\frac{10\dot{M}_{\rm r}R_{\rm S}}{\dot{M}_{\rm Edd}R}\right)^2}\right)^{-1}, \notag
\end{align}
with $f=1-\sqrt{R_{\rm ISCO}/R}$, and $R_{\rm ISCO}$ a function of black hole spin.

The existence of suitable emission or absorption lines for kinematic identification of a recoiling TDE is an open question.  There is widespread agreement in the literature on the existence of emission lines in the immediately unbound, photoionized stellar debris (Bogdanovic et al. 2004, Strubbe \& Quataert 2009, Strubbe \& Quataert 2010).  Some of these spectral features are potentially useful for distinguishing TDEs from supernovae and other transients (Kasen \& Ramirez-Ruiz 2010).  Unfortunately, the velocity spread in this debris, $\Delta v \approx (2\Delta \epsilon)^{1/2}$, can easily reach a large fraction of the speed of light, and details of the Doppler broadening will be determined by the inclination of the star's initial orbit around the black hole, among other unknown parameters.  A more promising candidate seems to be absorption lines formed when outflowing material processes continuum radiation from the accretion disk around the black hole (Strubbe \& Quataert 2010).  Most of these lines are in the UV part of the spectrum, although weaker hydrogen or helium lines may exist in optical bands, and the presence of a soft X-ray, power law tail (as has been observed in tidal flares detected by ROSAT and GALEX - see Komossa 2002, Gezari et al. 2008) could produce absorption lines between 1 and 10 keV as well (Strubbe \& Quataert 2010).  If the wind launching speed does not vary too much in time, these lines will be sufficiently narrow (with thermal broadening $\sim 30 \rm ~km~s^{-1}$) to make the velocity difference between the outflowing wind's photosphere and the host galaxy measurable.  However, the bulk outflow velocity may itself be quite large, in which case determination of the black hole velocity would require extremely precise wind modeling.  Alternatively, if $f_{\rm v} \ll 1$, a recoiling black hole's velocity could dominate the kinematic offset between the photosphere and the galaxy, and make wind launching speed variations small.  A limiting case of this regime is a different model for the super-Eddington phase of mass return (Loeb \& Ulmer 1997), in which radiation pressure from the disk isotropizes returning gas streams and supports a quasi-spherical cloud of disrupted matter.  In this model, measurement of the absorption offset from host galaxy lines would precisely identify which TDEs were caused by recoiling black holes.  Although the validity of either of these models is at the moment unclear, recent observations of TDEs found in SDSS data (van Velzen et al. 2010) may be more compatible with the predictions of Loeb \& Ulmer 1997.  Further validation of this model (or a low $f_{\rm v}$ version of SQ09) would indicate the feasibility of kinematic identification of recoil-induced TDEs.

For a recoiling SMBH, there are two sources of stars to tidally disrupt: unbound stars encountered in its passage through the host galaxy, and the small cloud of stars in orbits that remained bound during the recoil event.  Both of these sources are considered in the next subsections.  For simplicity, in the remainder of this paper we set $\eta=1, m_*=M_{\odot}$, and $r_*=R_{\odot}$ (a conservative assumption - see Tremaine \& Magorrian 1999 for how generalizing to a more realistic stellar mass function can increase tidal disruption rates by a factor $\sim 2$).

\subsection{Host Galaxy Structure}

Like many of the stationary SMBH tidal disruption rate papers (Ulmer \& Syer 1998, Tremaine \& Magorrian 1999, Wang \& Merritt 2003) we consider the Nuker surface brightness parametrization (Lauer et al. 1995) 
\begin{equation}
I(r)=I_{\rm b}2^{\frac{B - \Gamma}{\alpha}}\left(\frac{r}{r_{\rm b}}\right)^{-\Gamma}\left(1+\left(\frac{r}{r_{\rm b}}\right)^{\alpha}\right)^{-\frac{B - \Gamma}{\alpha}},
\label{Nuker}
\end{equation}
which was originally developed to model the surface brightnesses of nearby ellipticals and bulges resolved at the parsec level with the Hubble Space Telescope.  In Equation (\ref{Nuker}), $r_{\rm b}$ is the break radius at which the shallow inner power law of $I(r)\propto r^{-\Gamma}$ becomes the outer power law, $I(r)\propto r^{-B}$.  The strength of the break is determined by the dimensionless coefficient $\alpha$, and $I_{\rm b}$ is the surface brightness at the break radius.  The profile can be Abel inverted to yield a volume density profile with power law exponents of $\gamma \approx \Gamma+1$ and $\beta \approx B+1$ when $r\ll r_{\rm b}$ and $r\gg r_{\rm b}$, respectively.  The Nuker parametrization has many well-known surface brightness models as special cases (Byun et al. 1996).  

The Nuker galaxies are roughly divided into two categories based on the steepness of the surface brightness profile in the centermost regions: core galaxies and cusp galaxies.  It has been suggested (Merritt \& Milosavljevic 2005) that core galaxies are formed when the inspiral and merger of two SMBHs ejects stars from the host galaxy's center via 3-body interactions, scouring a core from the stellar profile - though subsequent star formation can rebuild the central parsecs into a cusp in the case of gas-rich (so-called wet) mergers (Kormendy et al. 2009, Blecha et al. 2010).  For this reason, we will take core galaxies as those which best represent the immediate post-recoil state of the stellar profile in gas-poor (so-called dry) mergers.  We will consider cusp galaxies also, as they are likely to be relevant in wet merger scenarios.  Most core galaxies in the Nuker sample tend to possess large central black hole masses $M_{\rm BH}$ ($\ga 10^8 M_{\odot}$, as determined by the $M_{\rm BH}-\sigma$ relation).  To examine a wider range of galaxy masses, we do not directly use observed samples of galaxies but rather create a simple mock catalog of galaxies in which we simulate the black hole trajectories.  Important scaling relations which we require for all galaxies in our mock catalog are the $M_{\rm BH}-\sigma_*$ relation (Tremaine et al. 2002),
\begin{equation}
\sigma_*=220{\rm km~s^{-1}}\left(\frac{M_{\rm BH}}{2\times10^8M_{\odot}}\right)^{0.249},			   
\end{equation}
and the bulge-black hole mass relation (Marconi \& Hunt 2003)
\begin{equation}
M_*=4.06\times10^{10}M_{\sun}\left(\frac{M_{\rm BH}}{10^8M_{\odot}}\right)^{1.04},
\end{equation}
with $M_*$ being total stellar mass and $\sigma_*$ the bulge velocity dispersion.  We also fit our galaxies to empirically determined calibrations of SMBH influence radii (Merritt et al. 2009):
\begin{equation}
r_{\rm infl}=35~{\rm pc}~\left(\frac{M_{\rm BH}}{10^8 M_{\odot}}\right)^{0.56},
\end{equation}
\begin{equation}
r_{\rm infl}=22~{\rm pc}~\left(\frac{M_{\rm BH}}{10^8 M_{\odot}}\right)^{0.55}.
\end{equation}
The first of these relations holds for core galaxies and the second for cusp galaxies.  Here the influence radius, $r_{\rm inf}$ is the radius within which there is a mass in stars equal to $2M_{\rm BH}$.

The most relevant aspect of the Nuker parametrization for our purposes is the broken power-law in the stellar density profile; as we shall see, the trajectories of kicked black holes are strongly affected by central density gradients.  While the most self-consistent way to proceed would be to numerically deproject scaled brightness profiles $I(r)$ into luminosity density profiles $j(r)$, we suspect this would fail to capture an important part of the physics of black hole recoil.  Initially radial orbits in spherically symmetric potentials make multiple passes through galactic center, whereas initially radial orbits in nonspherical potentials will generally receive torques that increase their pericenter distance.  This is of importance to this paper because orbital energy loss for recoiling black holes in dry mergers is dominated by stellar dynamical friction during passes through the densest central regions (Gualandris \& Merritt 2008).  Using a spherical potential would produce purely radial orbits, artificially shortening the wandering life time of the black hole, and decreasing the number of offset TDEs it can produce.  To avoid this problem, we use an ad hoc density-potential pair that is designed to approximate the most important dynamics of a Nuker profile but which is also easily generalizable to nonspherical geometries.  Specifically, we create a set of spheroidal isodensity surfaces

\begin{equation}
\rho_*(m)=
\begin{cases}
K_1m^{-\gamma} & m<m_{\rm b}\\
K_2m^{-\beta} & m_{\rm b} \le m < m_{\rm max} \\
0 & m\ge m_{\rm max}
\end{cases}
\label{rho}
\end{equation}

using the axisymmetric pseudoradius
\begin{equation}
m^2=\frac{r^2}{a^2}+\frac{z^2}{c^2}.
\end{equation}
In these equations $r$ and $z$ are standard cylindrical coordinates, and the isodensity spheroids have dimensionless axis ratios $a$ and $c$ ($a>c$).  The so far free parameters $\gamma$, $\beta$, and $m_{\rm b}$ are calibrated using the corresponding values in the spherical spatial deprojections of the most recent Nuker sample (Lauer et al. 2005).  None of those three parameters vary strongly with $M_*$ once the core/cusp dichotomy is accounted for, so we treat them as having Gaussian distributions about their measured means, with variance also calibrated off the Lauer et al. 2005 sample.  $K_1$ and $K_2$ are chosen so that the density profile is continuous at the break pseudoradius, $m_{\rm b}$, and $m_{\rm max}$ is selected in combination with $K_1, K_2$ to both normalize the total stellar mass, $M_*$, and to reproduce the correct value of $r_{\rm infl}$.  The final free parameter of our mock galaxy catalog is ellipticity $\epsilon = 1-c/a$.  Fortunately for our purposes, Lauer et al. (2005) measured isophote ellipticities for the galaxies in their sample, which we use to sample ellipticity parameter space (again, due to its lack of variation with $M_*$, we treat ellipticity as Gaussian-distributed).

The distinction between cores and cusps deserves more consideration, however.  Because SMBHs above $10^8~M_{\odot}$ consume stars whole rather than tidally disrupting them, the majority of galaxies of interest to us fall into the mass range represented primarily by cusps.  Although scaling relations for core galaxy parameters exist (Faber et al. 1997), it is not clear how reliably they can be extrapolated an order of magnitude below the smallest core galaxies in the samples which they are based on.  For our mock catalog, we consider the mergers of initially cuspy galaxies, but treat separately two different limiting scenarios:
\begin{itemize}  
\item In the first, the tendency of a binary SMBH inspiral to scour a core is outmatched by star formation, and a nuclear cusp is preserved.  In this pure cusp scenario, all free parameters \{$\gamma, \beta, r_{\rm b}, \epsilon$\} are calibrated off the sample of cusp galaxies.  We assume in this case that a fraction $f_{\rm g}$ of the baryonic mass of the bulge mass is in the form of a gas disk, which we describe in more detail in \S 2.4.
\item In the second, we consider a SMBH binary which is successful at scouring a core, either because its progenitor merger was dry or because free gas was consumed or expelled prior to the late phase of the hard binary's evolution.  For this case we calibrate the constant value of $\beta$ and the initial values $\gamma_0, r_{\rm b, 0}$ off the cusp sample, but calibrate $\epsilon$ and final $\gamma_{\rm f}$ off the core sample, and manually ``excavate" a mass deficit $\Delta M=2M_{\rm BH}$ to determine the final break radius $r_{\rm b, f}$.  The final value for the new, cored break radius is 
\begin{equation}
r_{\rm b, f}=\left(\frac{\frac{\Delta M}{4\pi K_1(1-e^2)^{1/2}}-r_{\rm b, 0}^{3-\gamma}\frac{\beta-\gamma_0}{(3-\beta)(3-\gamma_0)}}{r_{\rm b, 0}^{\beta-\gamma}\frac{\gamma_{\rm f}-\beta}{(3-\gamma_{\rm f})(3-\beta))}}\right)^{1/(3-\beta)}
\end{equation}
In practice, this tends to increase the break radius by a factor of a few.  Here we have used the ellipsoidal eccentricity $e=\sqrt{1-(c/a)^2}$.  In this second scenario, we set the remnant gas fraction $f_{\rm g}=0$.
\end{itemize}

With our galaxy models fully determined, we can then integrate our density profiles numerically to calculate other relevant quantities such as the stellar potential and forces.  Here we use the standard method of homoeoids (Binney \& Tremaine 2008, Section 2.5).  Due to the difficulty involved in even numerical calculation of a two-integral distribution function (Hunter \& Qian 1993), particularly when an analytical, closed-form potential is lacking (as is the case here) we are forced to compute the velocity dispersion in the spherical limit and generalize by substituting $m$ for $r$; this introduces modest inaccuracy into the dynamical friction and gravitational focusing formulae used in \S 2.4.  The spherical limit of this broken power-law profile has a velocity dispersion
\begin{align}
\sigma^2(r) &=2\pi G K_2(\frac{2r_{\rm b}^{3-\beta}}{r(3-\gamma)(1+\beta)}\label{sigma1} \\
&+\frac{r^{2-\beta}}{(3-\beta)(\beta-1)}-\frac{2r_{\rm b}^{3-\beta}}{r(3-\beta)(1+\beta)})\notag\\
\notag\end{align}
for $r>r_{\rm b}$, and
\begin{align}
\sigma^2(r) &=2\pi G K_1 r_{\rm b}^{2-2\gamma}r^{\gamma}(\frac{2}{(3-\gamma)(1+\beta)}\notag\\
&+\frac{1}{(3-\beta)(\beta-1)}-\frac{2}{(3-\beta)(1+\beta)}\label{sigma2} \\
&+\frac{1}{(3-\gamma)(1-\gamma)})-\frac{2\pi G K_1 r^{2-\gamma}}{(3-\gamma)(1-\gamma)}\notag\\
\notag\end{align}
for $r\le r_{\rm b}$.

The stellar bulge population dominates gravitational effects for low velocity kicks, but higher velocity kicks carry SMBHs into regions where the dark matter halo potential becomes important.  We model the density profile of dark matter using a standard NFW profile (Navarro et al. 1997), given by
\begin{equation}
\rho_{\rm NFW}=\frac{\rho_0}{(r/a_{\rm s})(1+r/a_{\rm s})^2}.
\end{equation}
Here the scale distance $a_{\rm s}$ and density factor $\rho_0$ are determined by assuming a concentration of 10 and truncating the NFW profile at $r_{200}$, the virial radius at which $\rho_{\rm NFW}=200\rho_{\rm c}$, with $\rho_{\rm c}$ the cosmological critical density.  We also normalize the total dark matter mass $M_{\rm tot}$ using the latest calibration of the $M_{\rm BH}$-$M_{\rm tot}$ relation (Bandara et al. 2009),
\begin{equation}
M_{\rm BH}=1.51\times10^{8}\left(\frac{M_{\rm tot}}{10^{13}M_{\odot}}\right)^{1.55}.
\end{equation}
Our assumption of spherical symmetry for the dark matter halo should be considered conservative, since nonspherical potentials extend the wandering time of the SMBH.

Our general strategy for galaxy modeling is to select a fiducial value of SMBH mass, a parameter set \{$\gamma, \beta, r_{\rm b}, \epsilon$\} informed by the Lauer et al. (2005) sample, and then to use the scaling relations described in this section to self-consistently find other parameters so that integrated quantities like potential or velocity dispersion can be numerically computed.  The fiducial values of $M_{\rm BH}$ we use are $10^6, 10^{6.5}, 10^7, 10^{7.5}$, and $10^8M_{\odot}$.  To parallel the SQ09 calculation we also use the black hole mass function inferred by Hopkins et al. (2007b).

In minor mergers, the dynamical friction timescale for satellite infall is expected to be greater than the Hubble time (Wetzel \& White 2009), meaning that we only need to consider the mass range of $q=0.1~-~1$.  Simulated merger rates for this mass ratio range in galaxies with stellar mass $M_*>M_{\rm min}$ have been matched (Hopkins et al. 2009) to the analytic fit
\begin{align}
\frac{dN_{\rm major}}{dt}=&0.04\left(1+\left(\frac{M_{\rm min}}{M_0}\right)^{0.8}\right)\\
\times &(1+z)^{\beta(M_{\rm min})}\rm ~Gyr^{-1} \notag, 
\end{align}
with $z$ the redshift of the galaxies, $M_0=2\times10^{10}M_{\odot}$, and 
\begin{equation}
\beta(M_{\rm min})=1.5-0.25\log\left(\frac{M_{\rm min}}{M_0}\right).
\end{equation}
We use this merger rate in combination with the volumetric SMBH mass function and the results of our orbit integrations (described below) to compute the total rate of events observable by LSST.  For all calculations in this paper we assume a standard $\Lambda$CDM cosmology with $\Omega_{\Lambda}=0.73$, $\Omega_{\rm M}=0.27$, and $H_{0}=71~{\rm km~s^{-1}~Mpc^{-1}}$ (Spergel et al. 2007).

\subsection{Interactions with the Galaxy}
To model the disruption rate of unbound stars, we evolve the SMBH's trajectory under the influence of gravity and dynamical friction through galaxies with the axisymmetric Nuker density profiles described above.  We use the fifth-order Dormand-Prince method (with an embedded fourth-order Runge-Kutta integrator for adaptive timestepping) described in Numerical Recipes, Chapter 17.  The effects of dynamical friction are approximated with the Chandrasekhar formula (Chandrasekhar 1943),
\begin{equation}
{\bf F}_{\rm df}=-I(M)\frac{4\pi\rho(GM_{\rm BH})^{2}}{\sigma^{2}}\frac{{\bf v}_{\rm BH}}{v_{\rm BH}},
\end{equation}
with $\sigma$ the local velocity dispersion and $\rho$ the local density of the medium causing the drag.  For a collisionless medium, such as a stellar population, 
\begin{equation}
I_{\rm dry}(M)=\frac{{\ln}(\Lambda)}{M^{2}}\left({\rm erf}\left(\frac{M}{\sqrt{2}}\right)-\sqrt{\frac{2}{\pi}}Me^{-M^{2}/2}\right)
\label{Idry}
\end{equation}
where $M=v_{\rm BH}/\sigma$ is the Mach number.  The Coulomb logarithm can be fit numerically (Escala et al. 2004), and for the case of SMBHs on radial orbits, is well fit by a value of ${\rm ln}\Lambda=2.5$ (Gualandris \& Merritt 2008).  Dynamical friction is the force which ultimately causes the kicked black hole to settle back into a near-stationary position in the center of its host galaxy, on timescales ranging from $10^6$ to $10^9$ years (Madau \& Quataert 2004, Blecha \& Loeb 2008).  The Chandrasekhar formula is derived assuming a uniform and infinite background of stars, and it is not immediately clear how appropriate that is for a steep density profile in galactic nuclei, or for a black hole massive enough to excite a response in the stellar population.  The applicability of the Chandrasekhar formula to bound, recoiling black holes has been considered before (Gualandris \& Merritt, 2008), and for appropriately chosen $\ln \Lambda$ it was found to be fairly accurate until the point when the mass interior to the black hole's apogalacticon is of order $M_{\rm BH}$.  After this, coherent oscillations develop in the stars interior to the black hole's trajectory, and dynamical friction is found to become dramatically less effective at removing the black hole's orbital energy.  We terminate our calculations at the onset of this orbital phase, both because our trajectory approximation would become quite inaccurate and also because TDEs caused by a slow-moving SMBH near the center of a galaxy would not be distinguishable from those caused by a stationary black hole.  During these calculations, we neglect the extra ``core scouring'' caused by black hole recoil (Gualandris \& Merritt 2008).  The stellar population in the galactic center responds to a moving SMBH by expanding, with a mass of stars equal to a few $M_{\rm BH}$ being displaced from the galactic center for kicks close to escape velocity (and the effect is reduced for slower ones).  Neglect of this effect likely reduces the SMBH wandering time and causes us to underestimate the total number of TDEs per galaxy merger, but probably not by much, as axisymmetry of the stellar potentials prevents the SMBHs from returning exactly to the center of their host galaxies where core scouring is most relevant.  We highlight that dynamical friction removes the most orbital energy during passages through the densest regions of the SMBH's trajectory.  Therefore trajectories with nonzero angular momentum (due to an axisymmetric potential) last longer before settling back into the galactic center than would center-crossing ones in spherical geometries.

In a dry merger it is sufficient to consider dynamical friction off stars and not gas.  This regime could also apply to wet mergers where the gas is used up in star formation (while the SMBH binary is stalled) or dispersed in binary quasar feedback.  We identify both of these scenarios with our ``excavated core'' galaxies.  If significant quantities of gas survive until the recoil phase of the merger, however, it is necessary to consider the effects of gas dynamical friction, which would apply more to our ``pure cusps.''  Previous work (Blecha et al. 2010) has indicated the effect of leftover gas is to decrease black hole wandering times, reducing the observable number of offset TDEs.  To quantify this effect, Equation (\ref{Idry}) still applies; we simply need to use gas density rather than stellar density for $\rho$, substitute a local sound speed $c_{\rm s}$ for $\sigma$, and modify the dimensionless parameter $I(M)$ (for gas, $M=v_{\rm BH}/c_{\rm s}$).  The new dimensionless functions are
\begin{equation}
I_{\rm subsonic}(M)=\frac{1}{2}\ln\left(\frac{1+M}{1-M}\right)-M
\end{equation}
in the subsonic regime, and 
\begin{equation}
I_{\rm supersonic}(M)=\frac{1}{2}\ln\left(1-\frac{1}{M^2}\right)+\ln\left(\frac{v_{\rm BH}t}{r_{\rm min}}\right)
\end{equation}
in the supersonic regime (Ostriker 1999).  However, these formulae have been shown to overestimate gas dynamical friction in the slightly supersonic regime, so we adopt the prescription of Escala et al. 2004 and use the Chandrasekhar formula for $I(M)$, with ${\rm ln}\Lambda=2.5$ for $M\ge 4.7$ and ${\rm ln}\Lambda=1.5$ for $M<0.8$.  We follow the prescriptions of Blecha \& Loeb (2008) and assume that most of the gas in the galaxy has settled into a disk, which we align with the oblate plane of the galaxy.  We employ a slightly less complicated version of their model, as only two of their four disk zones are relevant for our dynamical modeling: zones III and IV (zones I and II only exist in the presence of a central SMBH).  Zone III, the portion of the disk influenced by the SMBH potential before the recoil kick, is truncated on its inner edge at the kick radius,
\begin{equation}
r_{\rm k}=\frac{GM_{\rm BH}}{v_{\rm k}^{2}}, 
\end{equation}
and transitions to zone IV at $r=r_{\rm infl}$.  Zone IV is an exponential disk with scale $r_{\rm disk}$.  The disk surface density in zone III will be 
\begin{equation}
\Sigma_{\rm III}=\left(\frac{4}{\pi Q^2}\right)\left(\frac{\dot{M}^2_\alpha}{\alpha^2G}\right)^{1/3}r^{-1},
\label{SigmaIII}
\end{equation}
while in zone IV, the surface density is
\begin{equation}
\Sigma_{\rm IV}=\frac{M_{\rm disk}(>r_{\rm infl})}{2\pi r_{\rm disk}(r_{\rm infl}+r_{\rm disk})e^{-r_{\rm infl}/r_{\rm disk}}}e^{-r/r_{\rm disk}}.
\end{equation}
Here we take the viscosity parameter $\alpha=0.1$, and set the Toomre parameter $Q=1$ (Toomre 1964) under the assumption that star formation feedback roughly balances cooling, leaving the disk marginally stable.  The scale distance $r_{\rm disk}$ is found by requiring continuity between zones III and IV: $\Sigma_{\rm III}(r_{\rm infl})=\Sigma_{\rm IV}(r_{\rm infl})$.  The accretion rate $\dot{M}_{\alpha}$ can be found by mass normalization of equation (\ref{SigmaIII}) so that $M_{\rm III}=2f_{\rm gas}M_{\rm BH}$:
\begin{equation}
\dot{M}_{\alpha}=\left(\frac{f_{\rm gas}M_{\rm BH} Q^2}{4(r_{\rm infl}-r_{\rm k})}\right)^{3/2}\alpha G^{1/2}.
\end{equation}
Densities in both disk zones decay exponentially with height $z$, with scale height
\begin{equation}
h_{\rm III}=\frac{Q^2}{8}r
\end{equation}
taken from Blecha \& Loeb 2008.  In zone IV, the scale height
\begin{equation}
h_{\rm IV}=\frac{\dot{M}_{\alpha}\kappa_{\Omega}}{3\pi^2\alpha QG\Sigma^2_{\rm IV}}
\end{equation}
is solved for using the identities $\dot{M}_{\alpha}=3\pi \alpha c_{\rm s}h\Sigma$ and $c_{\rm s}=(G\dot{M}_{\alpha}/\alpha)^{1/3}$.  Here $\kappa_{\Omega}$ is the epicyclic frequency, and is calculated from the numerically integrated potentials of the isodensity shells in Equation (\ref{rho}).

In our models we consider values of $f_{\rm g}$ (gas as a fraction of total baryonic mass) of 0 and 0.3.  The latter value is taken as a conservative upper limit for remnant gas fraction at the time of black hole merger, as self-consistent hydrodynamical simulations (Mihos \& Hernquist 1996) have shown that $\ge 50$\% of the initial gas fraction, $f_{\rm g,i}$ in a merger is expelled or converted into stars by the time of black hole coalescence.  Observation indicates that $f_{\rm g, i} \lesssim 0.6$ for $M_{\rm BH}>10^6M_{\odot}$ at low redshift (Hopkins et al. 2009, Figure 7), so $f_{\rm g}=0.3$ is a conservative case, likely to result in SMBH orbits which decay somewhat faster and produce fewer offset tidal disruptions than in more general wet mergers with smaller $f_{\rm g}$.

At each point along the SMBH's trajectory we consider an instantaneous ``tidal disruption cylinder'' of length $v_{\rm BH}\Delta t$ and radius equal to the gravitationally focused tidal disruption radius.  This lets us simply calculate instantaneous tidal disruption rates along the trajectory, 
\begin{equation}
\dot{N}_{\rm u}=\rho_*v\pi r_{\rm t}^2\left(1+\frac{2GM_{\rm BH}}{r_{\rm t}v^2}\right)
\end{equation}
which can be integrated to get a time-averaged TDE rate, or $N_{\rm TDE}$, the total number of stars disrupted per recoil event.  Here $v=\sqrt{v_{\rm BH}^2+\sigma^2}$, with $\sigma$ given by Equations (\ref{sigma1}) and (\ref{sigma2}).

\subsection{Interactions with the Bound Cloud}
The initial size of the bound cloud is determined by the magnitude of the received kick, and can be approximated as encompassing all stars within $r_{\rm k}$.  The mass of the bound cloud is found by KM08 to be a fraction $f_b$ of the black hole mass, where
\begin{equation}
f_b=F(\gamma)\left(\frac{2GM_{\rm BH}}{r_{\rm infl}v_{\rm k}^2}\right)^{3-\gamma}.
\end{equation}
Here $\gamma$ is the same as in Equation (\ref{rho}), $r_{\rm infl}$ is the influence radius, the interior of which contains a mass in stars twice $M_{\rm BH}$, and $F(\gamma)=11.6\gamma^{-1.75}$.  For most cloud sizes the disruption rate of bound stars will be determined by resonant relaxation into the SMBH's empty loss cone, exponentially depleting the population of stars inside on a timescale $\tau\approx 3.6GM^2_{\rm BH}/(v^3_{\rm k}m_*)$ (KM08).  In practice, the e-folding time is at least an order of magnitude below $10^{10}$ years for most of the $10^6M_{\odot}$ and $10^{6.5}M_{\odot}$ black holes which escape from their host galaxies, strongly suppressing the averaged intergalactic TDE rate.

One exception to this picture is if $r_{\rm k}\sim r_{\rm infl}$; in this case nonresonant relaxation could become important, and resonant relaxation alone will significantly underestimate the TDE rate.  This regime is of minimal significance for this paper, however, since small kicks are likely to produce few spectrally and no spatially offset flares.  A more significant exception is for relatively low-mass SMBHs, which can reach an energy relaxation timescale in less than their wandering time.  Relaxation will eventually allow the cloud to expand in radius (O'Leary \& Loeb 2009), changing the time evolution of the tidal disruption rate from exponential depletion to $\propto t^{-3/2}$ (O'Leary \& Loeb 2011).  Therefore, we adopt KM08's prescription for resonant relaxation,
\begin{equation}
\dot{N}_b\approx C_{\rm RR}(\gamma)\frac{\ln \Lambda}{\ln(r_{\rm k}/r_{\rm t})}\frac{v_{\rm k}}{r_{\rm k}}f_{\rm b}e^{-t/\tau}
\end{equation}
when $t<t_{\rm r}$, but transition to $\dot{N}_{\rm b} \propto t^{-3/2}$ at later times.  The energy relaxation timescale $t_{\rm r}$ is taken to be (O'Leary \& Loeb 2011)
\begin{equation}
t_{\rm r}=10^9 ~{\rm yrs} \left(\frac{M_{\rm BH}}{10^5M_{\odot}}\right)^{5/4}\left(\frac{r_{\rm k}}{r_{\rm infl}}\right)^{1/4}.
\end{equation}
This power law disruption rate is only relevant for $M_{\rm BH}<10^{6.5}M_{\odot}$, but for lower mass SMBHs we transition to power law depletion after an energy relaxation time.  For both scenarios, the initial disruption rate is
\begin{align}
\dot{N}_b\approx 1.5\times &10^{-6}\left(\frac{M_{\rm BH}}{10^7M_{\odot}}\right)\left(\frac{r_{\rm infl}}{10~{\rm pc}}\right)^{-2} \\
&\times \left(\frac{v_{\rm k}}{10^3~\rm km~s^{-1}}\right)^{-1}~\rm yr^{-1}. \notag
\end{align}
One uncertainty is the resonant relaxation coefficient $C_{\rm RR}$, found by KM08 to have a value of 0.14 for $\gamma = 1$.  Since the spatial power law exponents for core galaxies are close to 1, we adopt this value, though it is less well motivated for cuspier galaxies. 

We also consider growth of the bound cloud by capture of members of binary star systems.  This three-body interaction is treated in the same way as tidal disruption of unbound stars, except instead of a stellar tidal disruption radius we use an ``orbital tidal disruption radius'', given by 
\begin{equation}
r_{\rm t,o}=a_{\rm bin}\left(\frac{M_{\rm BH}}{2m_{*}}\right)^{1/3},  
\end{equation}
where $a_{\rm bin}$ is the binary semimajor axis.  While one member of the binary is ejected at high velocities (Hills 1988), the other is bound to the black hole, with apoapsis $r_{\rm max}$ given by
\begin{equation}
r_{\rm max} \approx \frac{GM_{\rm BH}}{v_{\rm eject}^2}\left(\frac{m_*}{M_{\rm BH}}\right)^{1/6}\left(\frac{a}{0.1\rm ~AU}\right)^{1/2}
\end{equation}
(Yu \& Tremaine 2003), with $v_{\rm eject} \approx 145 \rm km~s^{-1}$ (Hills 1988).  To calculate the rate of these captures, we assume $\rm\ddot{O}$pik's Law (\"Opik 1924), a flat distribution of binary semimajor axes $a$ in units of $\log(a)$, between $a_{min}$ and $a_{max}$.  Following Vereshchagin et al. (1988) and Kouwenhoven et al. (2007) we adopt $a_{min}=5R_{\sun}$ and $a_{max}=5\times 10^6 R_{\sun}$.  We only consider captures with $r_{\rm max} < r_{\rm t, c}$, with the cloud's tidal radius conservatively given by $r_{\rm t, c}=r_{\rm infl}$.  This refill mechanism is in principle capable of counterbalancing losses due to tidal disruption and evaporation from the cloud.  Without a refill source, resonant relaxation into the loss cone will normally cause the population of the bound cloud to evolve due to $\dot{N}_{\rm TDE}\propto N$, leading to a population (and TDE rate) depleted exponentially in time.  We can roughly see the effect of stellar capture into the bound cloud if we assume 
\begin{equation}
\dot{N}=-kN+m,
\end{equation}
with $k$ the average frequency with which bound stars evaporate or are scattered into the loss cone and $m$ a time-averaged capture rate.  This differential equation has the solution
\begin{equation}
N(t)=N(0)e^{-kt}+\frac{m}{k}(1-e^{-kt}).
\end{equation}
By itself, the resonant relaxation process will deplete the bound cloud, but 3-body capture allows the number of stars in the cloud to asymptotically approach a nonzero value.  If the time-averaged binary capture rate is high enough (i.e. if $m/k > N(0)$) the size of the cloud would even grow over time.  The importance of this effect is determined for each galaxy/kick velocity pair.

A final consideration is stability of the bound cloud to perturbations.  Analytically, it seems unlikely that interactions with unbound stars will eject significant numbers of bound stars from the cloud: if the cloud stars are bound to the black hole with typical energy $E_{\rm bind} \sim -v^2_{\rm k} m_*$, and during encounters with unbound stars a change in energy $\Delta E \sim Gm_*^2/r_{\rm p}$ is available (where $r_{\rm p}$ is the closest approach of the two stars), encounters must be within $r<r_{\rm p}\sim r_{\rm k}(m_*/M_{\rm BH})$.  For a $10^7 M_{\odot}$ black hole on typical trajectories, this works out to at most $\sim 1$ unbound stars making close enough approaches to eject a bound star during the SMBH's passages through the bulge.

\subsection{Observability of Recoil-Induced TDEs}
To translate the total recoil-induced TDE rate into a rate of {\it identifiably} recoil-induced TDEs, it is necessary to consider observational constraints.  LSST's rapid cadence, high sensitivity and thorough sky coverage make it an ideal survey to detect disruption flares - as mentioned in \S2.2, LSST could detect up to thousands of TDEs per year.  LSST's limiting g-band magnitude is 25 (LSST Science Collaborations 2009); because of LSST's short cadence we assume any flares brighter than that will be detected.  The detectability of a spatial offset will depend on how well the TDE centroid can be distinguished from the host galaxy centroid after photometric frame subtraction.  For LSST the expected differential astrometric precision will be $\sim 0.7''/\rm SNR$ (LSST Science Collaborations 2009).  Using the LSST Science Manual's prescription for ${\rm SNR}^{-1}=\sigma_{\rm tot}=\sqrt{\sigma_{\rm sys}^2+\sigma_{\rm rand}^2}$, we infer astrometric precision by calculating the signal to noise ratio for each event in our sample.  We also calculate the rate of spectroscopically identifiable flares associated with a recoiling SMBH.  Although UV spectroscopy would be ideal, soft X-ray spectrometers - SXS, for example, on the planned ASTRO-H mission (Takahashi et al. 2010), expected to be operating contemporarily with LSST - should be able to identify the absorption lines discussed in \S2.2, if they exist with sufficient equivalent width.  To investigate this possibility, we consider a fiducial case of absorption lines at 10 keV, observed by SXS followup with an energy resolution of 7 eV.  If the outflowing wind can be accurately modeled, these lines would allow black hole velocities down to $\sim 200~\rm km~s^{-1}$ to be spectrally resolvable.  As mentioned before, it is not clear that the super-Eddington phase of accretion will produce winds in which a $\sim 200~\rm km~s^{-1}$ offset is detectable, so predictions of kinematic offsets should be regarded as somewhat hypothetical.  Because spatial and kinematic offsets are angle-dependent, we average the observable quantities over all inclination angles for the host galaxy.

\section{TDE Rate}
Using the potentials and frictional forces described above, we integrate the trajectories of five different black hole masses at nine different kick velocities and seven inclination angles in galaxies with eighteen different possible permutations of mass-independent structural parameters, for a total of $11340$ runs (the final factor of two comes from wet vs dry mergers).  During preliminary test runs, a very weak dependence of the wandering time on $\beta$ and $\gamma$ was apparent (once variation in $\beta$ and $\gamma$ due to the core/cusp dichotomy is allowed for), so we set those quantities equal to their average values.  Among the remaining structural parameters, we only varied $\epsilon$ and $r_{\rm b, 0}$.

We terminate our trajectory calculations after a Hubble time, if the black hole has left the stellar bulge (and its attendant sources of friction) with escape velocity, or upon the onset of the ``Phase II'' orbital oscillations of Gualandris \& Merritt (2008), discussed in \S 2.4.

To calculate the total observable rate of TDEs due to recoiling black holes, $\Gamma$, we use a modified version of Equation (31) in SQ09.  Specifically,
\begin{align}
\frac{{\rm d}\Gamma}{{\rm dln}M_{\rm BH}}=\int_{r_{\rm ISCO}}^{r_{\rm t}} \int_{0}^{{\rm d}_{\rm max}(r_{\rm p})} \! &4\pi r^2 f_{\rm sky} \frac{{\rm d}n}{{\rm dln}M_{\rm BH}}\label{dGamma} \\
\times \frac{{\rm d}\gamma(r, r_{\rm p})}{{\rm dln}r_{\rm p}} {\rm d}r {\rm dln}r_{\rm p} \notag
\end{align}
Here $d_{\rm max}(r_{\rm p})$ represents the maximum comoving distance a TDE flare with given pericenter $r_{\rm p}$ can be seen at, using a 25 AB g-band magnitude limit.  LSST will detect flares at cosmological distances, so it is necessary to employ a K-correction, which has a modest impact on $d_{\rm max}$.  The rate $\gamma(r, r_{\rm p})$ is integrated over inclination/azimuth angles and galaxy properties, and is the rate at which either TDE flares are produced at a distance $r$ by TDEs with $r_{\rm p}$.  Further cuts are added to the integrand to calculate the rate at which observably spatially offset TDE flares, or observably kinematically offset TDE flares are produced, using the criteria described in \S 2.6 (with an average over azimuthal angles to account for projection effects).  In our average over galaxy properties, we give $M_{\rm BH}$-dependent weights to the ``pure cusp'' and ``excavated core'' scenarios from \S 2.3.  These weights are determined by binning the the Lauer et al. 2005 galaxy sample and computing the fraction of cusps and cores in each mass bin (with the small minority of intermediate cases taken as 50\% core, 50\% cusp).  

To calculate $d_{\rm max}$ for super-Eddington flares we use Equations (\ref{RPhoto}), (\ref{TPhoto}), and (\ref{tEdge}), while to do the same for disk emission we use Equation (\ref{diskEmit}).  For simplicity, we neglect the less important emission from photoionized, unbound disruption debris, noting that this is a conservative approximation.  For disruptions from the bound cloud, resonant relaxation slowly diffuses stars across the loss cone in phase space, meaning that nearly all bound TDEs will have $r_{\rm p}\approx r_{\rm t}$.  Unbound stars will have a wider variety of $r_{\rm p}$, but the geometry of gravitational focusing will bias them towards $r_{\rm p}\approx r_{\rm t}$ as well.  For these reasons we simplify Equation (\ref{dGamma}) by taking $r_{\rm p}=r_{\rm t}.$  This approximation produces slightly more disk emission (due to physically larger disks), and significantly less luminous super-Eddington flares, than the flat distribution of TDEs across ${\rm ln}r_{\rm p}$ assumed in SQ09.  Consequently, our results show a much less pronounced difference in the observable TDE rate between the disk emission and super-Eddington outflows cases.  

We then interpolate the results of these trajectory calculations over three different black hole physics scenarios, as discussed in \S 2.1.  In the first scenario, a lack of free gas during the SMBH inspiral leaves the spin vectors of the SMBH binary randomly aligned with each other, producing a top-heavy kick distribution and a high average value of dimensionless spin ($a=0.73$).  The other two scenarios involve wet mergers with warmer and cooler gas, producing spin vectors aligned to within $30$ and $10$ degrees, respectively, and remnant mean spins of $a=0.88$ and $a=0.90$.  Because the disruption of stars by $10^8 M_{\odot}$ black holes is so sensitive to $a$ ($a>0.92$ required), we bracket these fiducial assumptions about remnant spin with $a=0$ and $a=0.95$ cases.  We also consider two different cases of tidal disruption physics; one in which the super-Eddington mass outflows proposed by SQ09 exist (for simplicity we take their canonical case of $f_{\rm v}=1$ and $f_{\rm out}=0.1$), and the other in which they do not.  In the latter, optical emission is limited to the Rayleigh-Jeans tail of the newly-formed accretion disk.

\section{Results and Discussion}

\begin{figure}
\includegraphics[width=84mm]{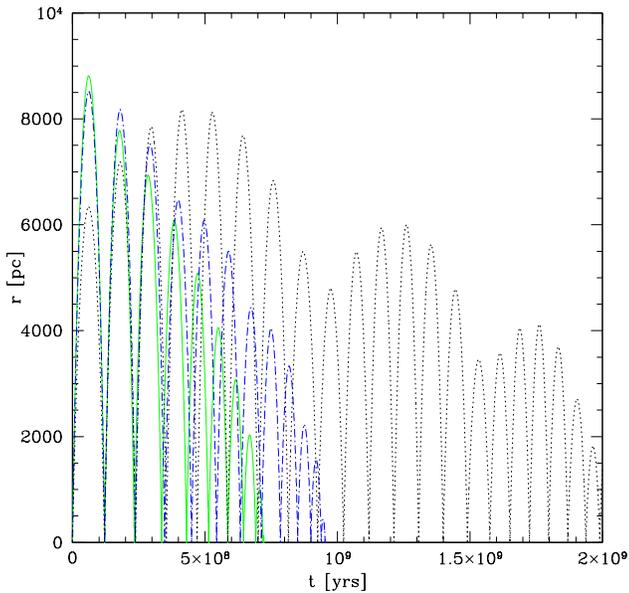}
\caption{The importance of nonsphericity on SMBH trajectories.  All the lines represent trajectories of $10^7~M_{\odot}$ SMBHs kicked at 400 $\rm km~s^{-1}$ in a core galaxy.  The green solid line is the course of a SMBH kicked in the plane of the galaxy's two semimajor axes, while the blue dot-dashed line represents a SMBH kicked 15 degrees above the plane, and the black dotted line a SMBH kicked 45 degrees above the plane.}
\label{radialTraj}
\end{figure}

\begin{figure}
\includegraphics[width=84mm]{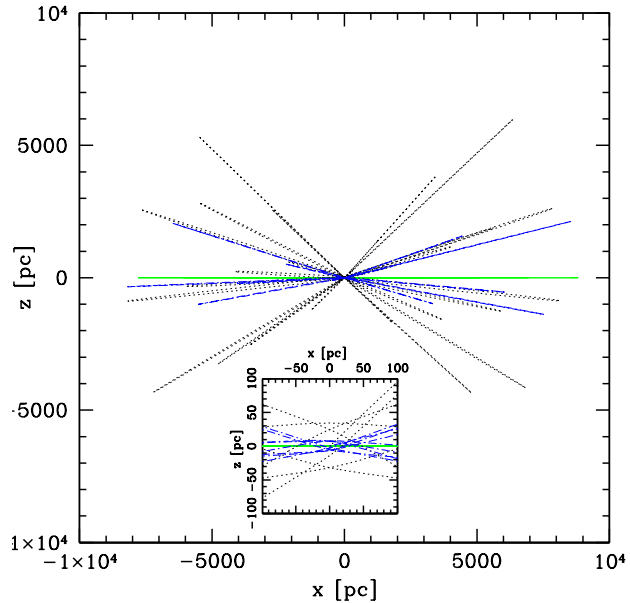}
\caption{The same black holes as in Figure 1, now viewed in two dimensions.  The inset plot zooms in on the central 100 parsecs to highlight the lack of center crossings for SMBHs ejected at nonzero inclination angles.}
\label{2DTraj}
\end{figure}

Figures \ref{radialTraj} and \ref{2DTraj} illustrate the importance of nonspherical potentials for the lifetimes of wandering black holes.  In the axisymmetric stellar potential that we employ, stars initially on radial orbits will quickly acquire angular momentum unless they lie on a principal axis of the stellar ellipsoid or in the equatorial plane.  The latter is true of the $0^{\circ}$, green orbit, which is seen in Figure 1 to decay somewhat faster than the blue, $15^{\circ}$ orbit and dramatically faster than the $45^{\circ}$, black orbit.  For the dry mergers illustrated here, the variation in decay time is due entirely to differences in stellar dynamical friction, which is the strongest at orbital pericenter.  In the wet merger scenarios we considered, recoils in the plane of the gas disk are very quickly damped out, but axisymmetry in the stellar potential still affects orbital lifetimes for other inclination angles.  Figure \ref{2DTraj} illustrates the torques that act on orbits out of the equatorial plane, and in the inset we can see that those torqued orbits are able to avoid close pericenter passages, explaining their longevity.

\begin{figure}
\includegraphics[width=84mm]{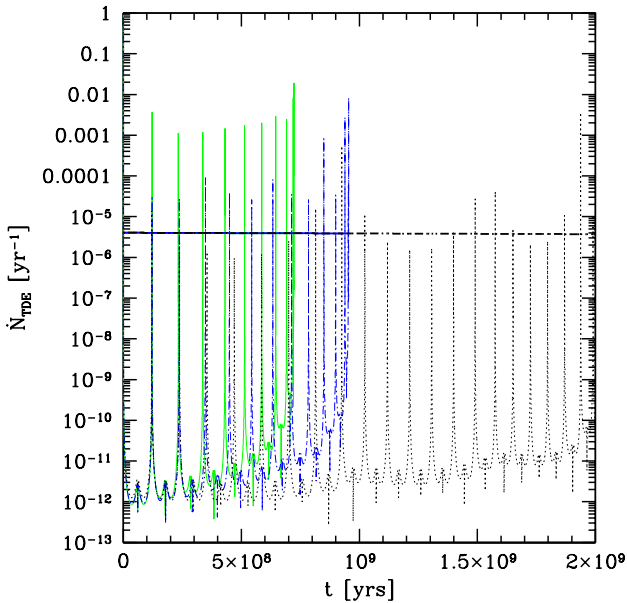}
\caption{The number of stars, $\dot{N}_{\rm TDE}$, disrupted per year for the SMBHs in previous figures.  The thick lines refer to disruptions from the bound cloud, while the thin lines refer to disruptions of unbound stars.  Colors and line types represent the same black holes as in Figure \ref{radialTraj}.}
\label{TDERate}
\end{figure}

\begin{figure}
\includegraphics[width=84mm]{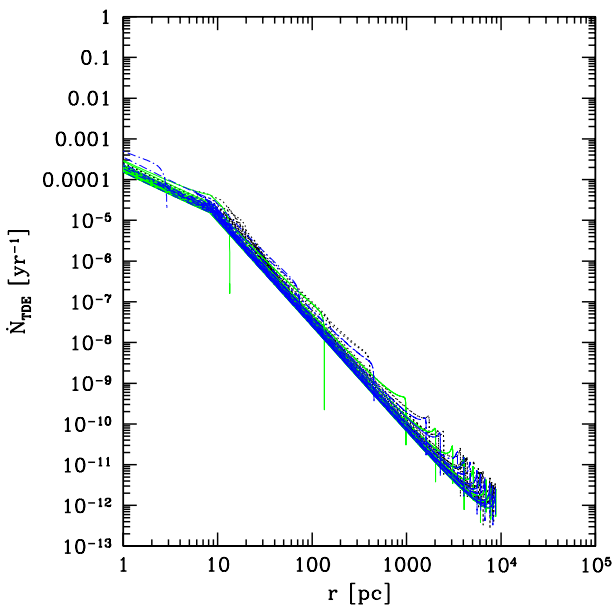}
\caption{The rate of unbound stellar disruption, $\dot{N}_{\rm TDE}$, as a function of radius for the SMBHs in previous figures.}
\label{TDERateRadial}
\end{figure}

The large density variations across the SMBH host galaxies can be seen in the unbound stellar disruption rate, displayed in Figure \ref{TDERate}, which spans many orders of magnitude.  The spikes are located at passages through the galactic center, and their increasing magnitude with each cycle arises from the inverse relation between gravitationally focused cross section and velocity.  The thick lines, representing the disruption rate of bound stars, are much more constant in time, although for lower-mass SMBHs these become observably depleted, as discussed earlier.  In Figure \ref{TDERateRadial}, we plot the unbound stellar disruption rate versus radial distance, and can clearly see the break in the stellar density profile near $7$ pc.  Scatter in Figure \ref{TDERateRadial} is due to the combination of different velocities and different inclination angles during pericenter passage.  The binary capture effect hypothesized in \S 2.5 was seen at low levels but found not to contribute substantially to bound cloud sizes or disruption rates.

The total number of disruptions per merger, $N_{\rm TDE}$, was found to be fairly insensitive to the power-law slopes $\gamma$ and $\beta$, but $1\sigma$ changes in $r_{\rm b}$ or $\epsilon$ can change $N_{\rm TDE}$ by a factor of a few.  The sensitivity to $r_{\rm b}$ really reflects a sensitivity to the ratio $v_{\rm k}/v_{\rm esc}$, as the wandering time can jump by $\sim 1-2$ orders of magnitude when $v_{\rm k}/v_{\rm esc}$ rises above a value $\sim 0.5-0.6$ (where $v_{\rm esc}$ here is the escape velocity of the stellar bulge).  This sensitivity to kick velocity is due to the decreased effectiveness of dynamical friction once the SMBH begins passing through the galactic center at high velocities, giving the black hole more time to disrupt bound cloud stars at an observably offset distance. 

\begin{figure*}
\includegraphics[width=180mm]{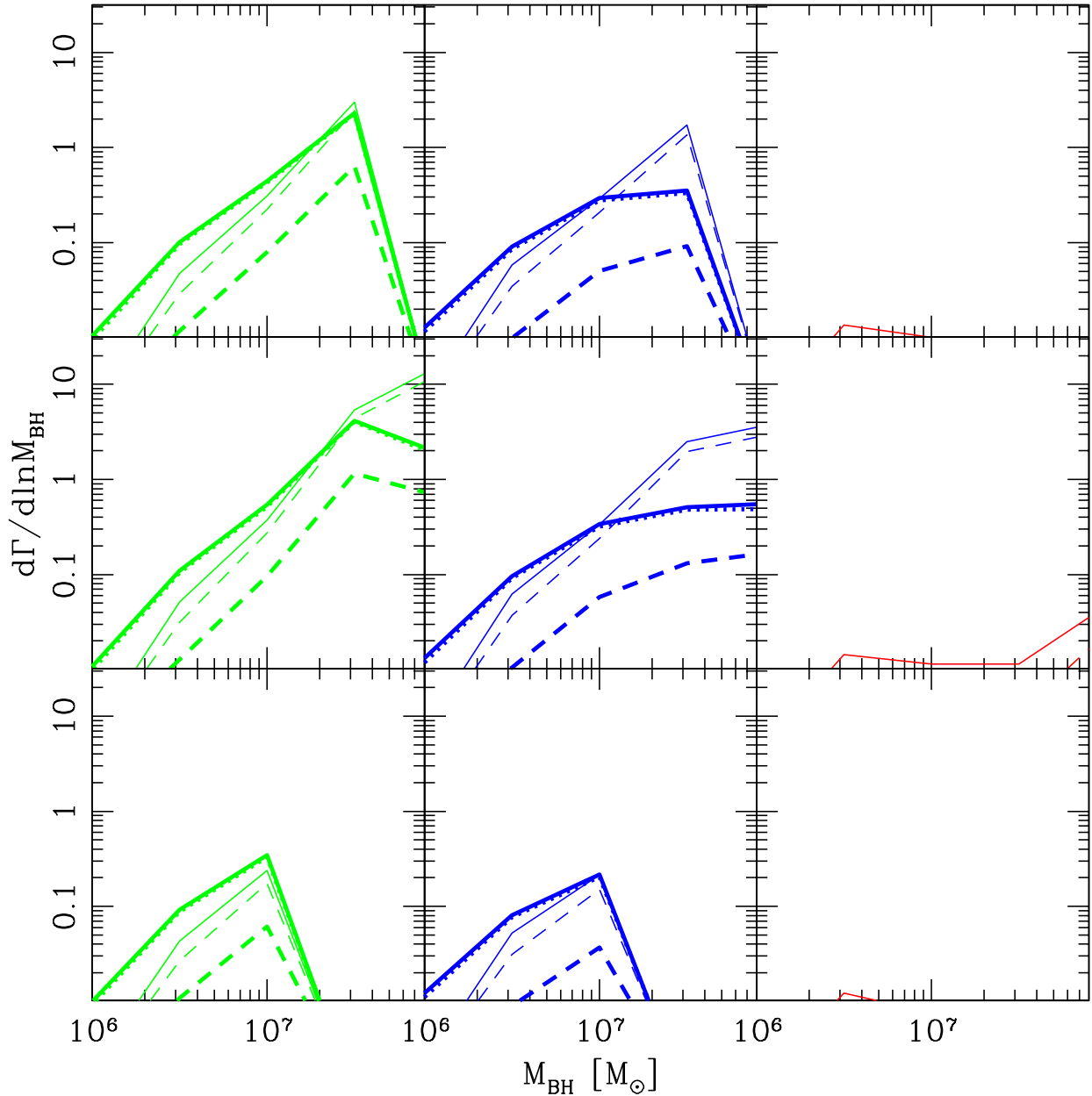}
\caption{Mass dependence of total TDE rates, $\Gamma$, for a variety of kick distributions and mean spins.  In the left (green) column, we show the unaligned spin case; in the middle (blue), spins aligned to within 30 degrees; and the right (red), spins aligned to within 10 degrees.  The top row has fiducial final spin amplitudes, while the middle has $a=0.95$ and the bottom row has $a=0.0$.  Within each frame, the thick lines represent disruptions of bound stars while the thin lines represent disruptions of unbound stars.  The solid lines represent total number of disruptions, while the dotted lines represent disruptions with an observable spatial offset and the dashed lines represent disruptions with an observable kinematic offset.  In this plot only disk emission is considered.}
\label{mosaic}
\end{figure*}

\begin{figure*}
\includegraphics[width=180mm]{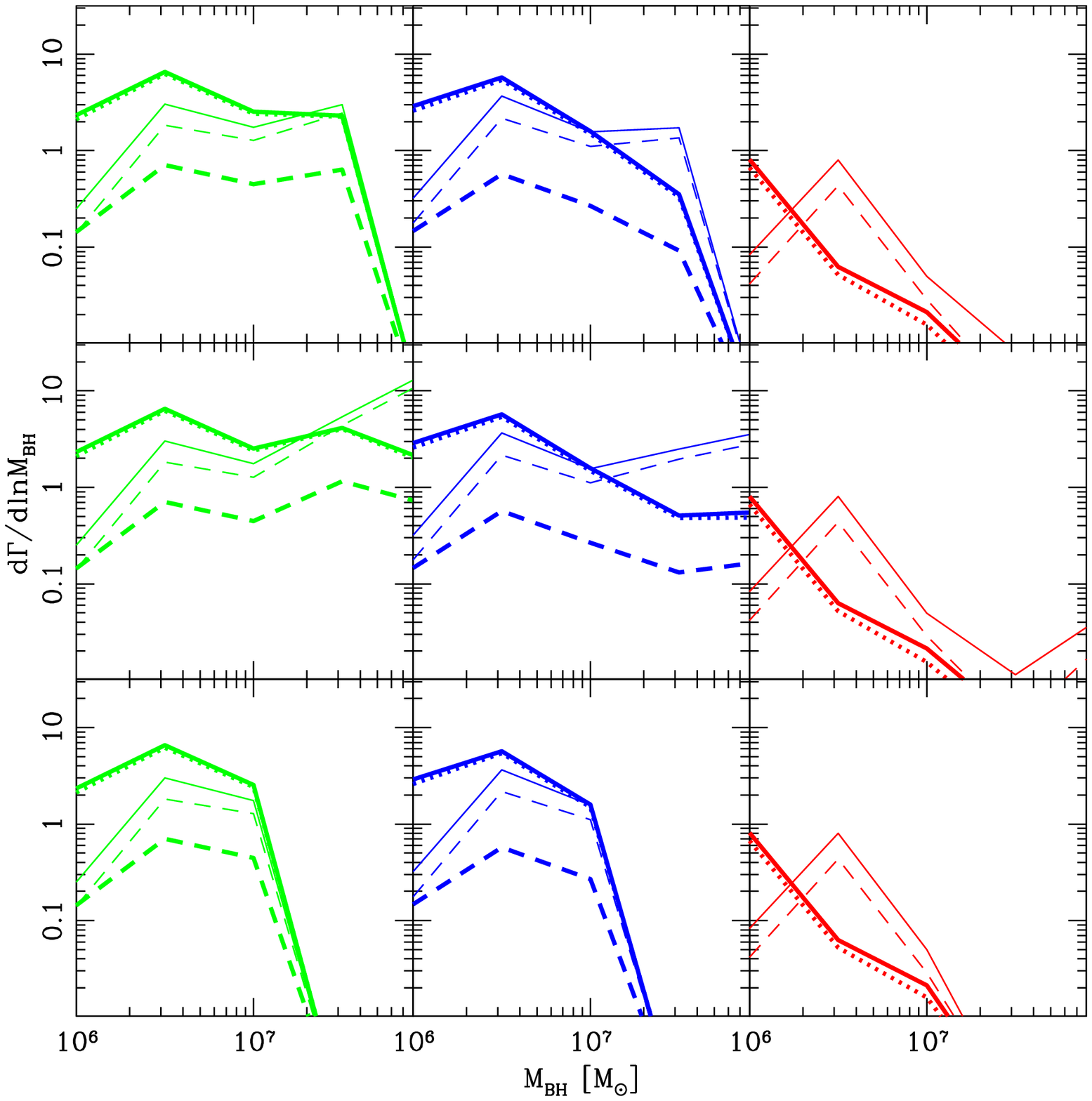}
\caption{The same as the previous figure, but assuming the existence of super-Eddington outflows.}
\label{mosaicSE}
\end{figure*}

Figures \ref{mosaic} and \ref{mosaicSE} illustrates how our results vary with assumptions about the kick velocity distribution, final spin amplitudes of the SMBHs, and existence of super-Eddington outflows.  In both figures, bound cloud disruptions are represented as thick lines and unbound stellar disruptions as thin lines.  The total number of disruptions is shown as a solid line, while those with an observable spatial offset are shown with a dotted line, and those with an observable kinematic offset are shown with a dashed line.  Unless otherwise noted, discussion of TDE rates in this section refers only to SMBHs which remain bound to their host galaxy.

Figure \ref{mosaic} displays ${\rm d}\Gamma/{\rm dln}M_{\rm BH}$, the number of TDEs observed by LSST per year per logarithmic black hole mass, for our models without super-Eddington emission.   Both the unaligned ($<180^{\circ}$) and moderately aligned ($<30^{\circ}$) progenitor spin models produce interesting values of $\Gamma$.  Both bound and unbound disruption rates are dominated by the highest black hole mass permitted by its spin amplitude to disrupt stars; for the first row (fiducial $a$ values), this corresponds to $10^{7.5}M_{\odot}$, while for the second row ($a=0.95$) it is $10^8M_{\odot}$ and for the third row ($a=0.0$) it is $10^7M_{\odot}$.  Almost all bound cloud disruptions have an observable spatial offset, while unbound disruptions never have an observable offset.  On the other hand, a higher fraction of unbound disruptions possess an observable kinematic offset relative to the bound cloud disruptions.  Both these correlations are easily explainable: due to high orbital eccentricity, the SMBHs in our sample spend the majority of their time far from the galactic nucleus, so most bound cloud disruptions occur with a large physical offset and low velocity.   At the same time, virtually all (see Figure \ref{TDERate}) unbound disruptions occur during perigalacticon, where the SMBHs move at their highest velocities.  The highly aligned ($<10^{\circ}$) progenitor spin model produces a negligible number of disruptions; high kick velocities are suppressed, and SMBHs escape the galactic nucleus too infrequently to disrupt significant numbers of stars.

In Figure \ref{mosaicSE}, we display ${\rm d}\Gamma/{\rm dln}M_{\rm BH}$ for models with super-Eddington flares.  The results are similar to those in Figure \ref{mosaic}, although ${\rm d}\Gamma/{\rm dln}M_{\rm BH}$ is everywhere greater than or equal to its values in the previous figure.  Two special points of contrast are the large increase in observable disruptions at the low end of the SMBH mass function, and the (corresponding) increase in disruptions for the highly aligned progenitor spin model.  The addition of super-Eddington flares has, as expected, little effect on values of ${\rm d}\Gamma/{\rm dln}M_{\rm BH}$ above $10^7M_{\odot}$, but disruption flares become dramatically more visible for $10^6M_{\odot}$ and $10^{6.5}M_{\odot}$ SMBHs.

\begin{figure}
\includegraphics[width=84mm]{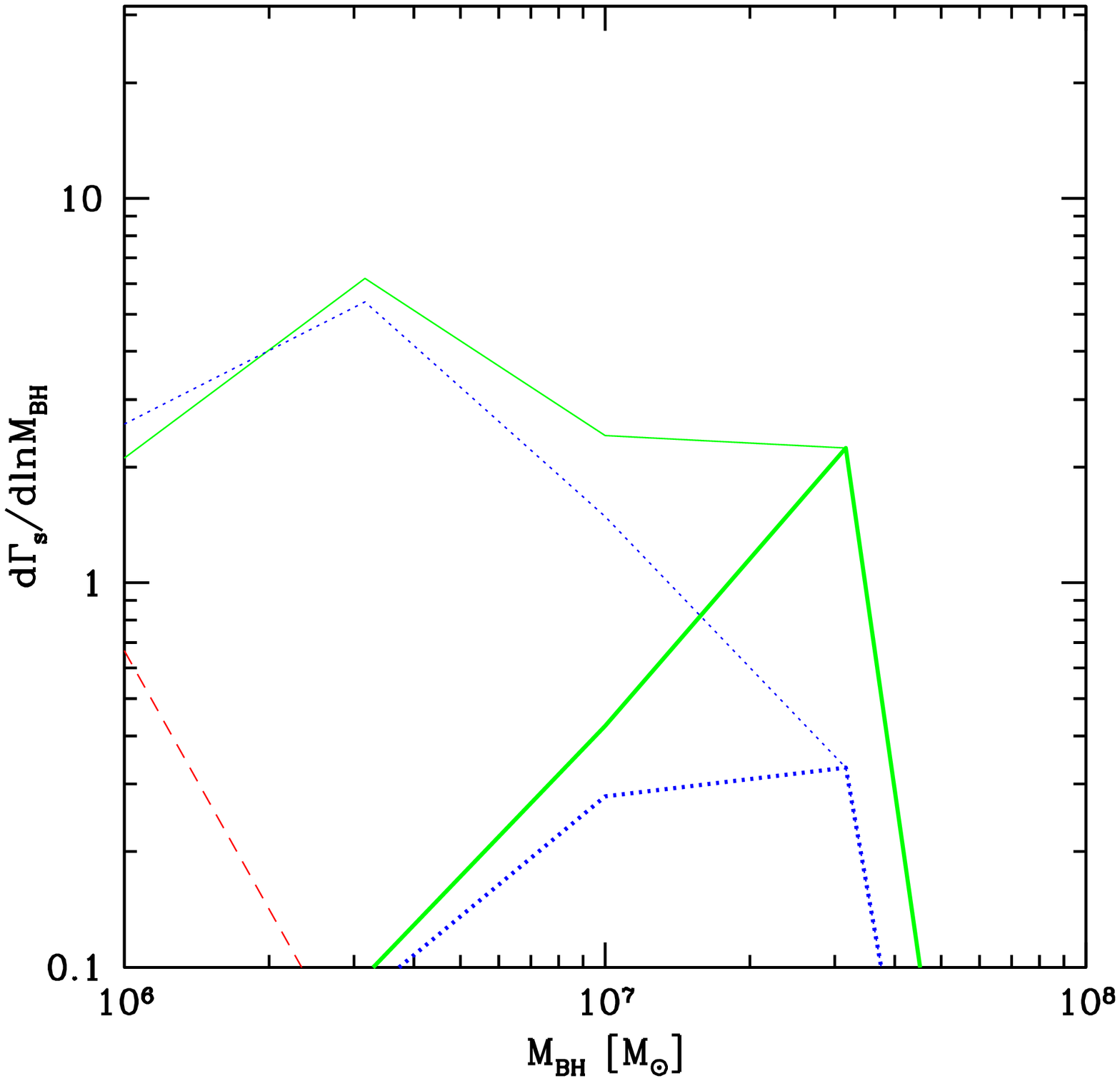}
\caption{Mass dependence of the total, galaxy- and velocity-averaged rate of spatially offset TDEs, $\Gamma_{\rm s}$.  $180^{\circ}$, $30^{\circ}$, and $10^{\circ}$ progenitor spin alignment correspond to solid green lines, dotted blue lines, and dashed red lines, respectively.  Thick lines represent disk emission only, while the thin lines correspond to disk plus super-Eddington outflows.}
\label{TotalGamma}
\end{figure}

Figure \ref{TotalGamma} displays ${\rm d}\Gamma_{\rm s}/{\rm dln}M_{\rm BH}$, the mass dependence of the total observable (spatially offset) rate $\Gamma_{\rm s}$, integrated over all kick velocities and all galaxies in our mock catalog, and given fiducial spin values.  When we integrate over black hole mass, we find that two of our kick velocity distributions produce a robustly observable ($\sim 10$) number of disruptions per year assuming super-Eddington flares, while the third produces a marginal number of TDEs, of order unity per year.  Likewise, progenitor spin distributions aligned to within $180^{\circ}$ or within $30^{\circ}$ produce $\sim 1$ flare per year with an observable spatial offset if we are only able to observe disk emission.  Higher-mass SMBHs contribute the most to observable disk flares, due to the lower temperatures and higher optical luminosities of their disks, while super-Eddington accretion flares are dominated by the lower-mass end of the SMBH distribution.  Although the rate enhancement from inclusion of super-Eddington outflows is almost a factor of $10$, this is considerably lower than the comparable factor in SQ09.  The reason for this disparity is that the brightest super-Eddington outflows correspond to the deepest plunges (lowest $r_{\rm p}$) into the tidal disruption region.  SQ09 considered a logarithmically flat distribution of $r_{\rm p}$, while we took a constant $r_{\rm p}=r_{\rm t}$, for the reasons explained in \S 3.

A variety of observable TDE rates are displayed in Table 1.  These numbers have been integrated over galaxy type, kick velocity distribution, inclination angle, and black hole mass function, and indicate that the ultimate observability of recoil-induced TDEs will depend strongly on both the existence of super-Eddington flares, and the average distribution of pre-merger spin alignments.  In this table, $\Gamma$ values for fiducial SMBH spins are shown, with the $a=0$ and $a=0.95$ cases appearing as lower and upper limits in parentheses.  It is only in the case where super-Eddington flares do not exist {\it and} substantial progenitor spin alignment occurs where we expect LSST to observe negligible numbers of spatially offset TDEs per year.  We note that if the progenitor spins are unaligned, or even aligned with scatter $\ge 30^{\circ}$, the tidal disruption rate from recoiling black holes is almost $1\%$ of the total TDE rate for all galaxies.  For most of our models, the number of kinematically offset TDEs, $\Gamma_{\rm k}$, is comparable to $\Gamma_{\rm s}$, although we note again that the theoretical basis for expecting appropriate absorption lines in a super-Eddington outflow is less secure than for a simple spatial offset.  We have also included in Table 1 the rates of spatially and kinematically offset TDEs for SMBHs which escape their host galaxy altogether, labeling these as $\Gamma_{\rm s, esc}$ and $\Gamma_{\rm k, esc}$.  Only in the case of unaligned spins and super-Eddington outflows are $\Gamma_{\rm s, esc}\sim \Gamma_{\rm s}$ and $\Gamma_{\rm s, esc}\sim \Gamma_{\rm k}$; in all other scenarios the number of observable TDEs due to ejected SMBHs is at least a factor of 7 smaller than the number due to bound SMBHs.

\begin{figure}
\includegraphics[width=84mm]{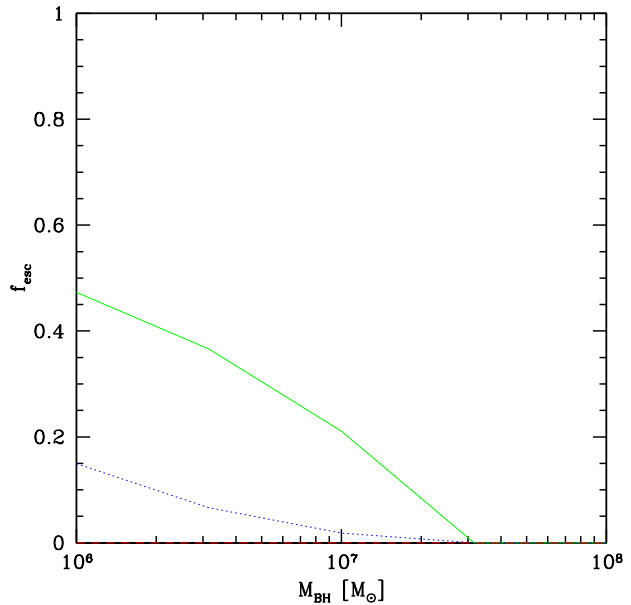}
\caption{Fraction of recoiled SMBHs which escape into intergalactic space, as a function of black hole mass, for the three different kick velocity distributions.  As in the previous figure, the green solid line represents the 180 degree alignment distribution, the dotted blue the 30 degree, and the dashed red (not visible; a negligible fraction of SMBHs from this distribution escaped their host galaxies) the 10 degree.}
\label{EscapeFraction}
\end{figure}

For most of the models we have considered, a large majority of the TDEs associated with recoiling SMBHs occur for black holes bound to their host galaxy.  This is due to two factors: both the relatively low fraction of SMBHs recoiled at escape velocity (see Figure \ref{EscapeFraction} for a plot of the SMBH escape fraction, $f_{\rm esc}$), and the smaller, more rapidly decaying bound clouds of those low-mass SMBHs which do escape.  This highlights the importance of searching for SMBHs bound to the bulge or halo of their host galaxy; although the intergalactic TDEs of the KM08 scenario offer a cleaner signal, they are intrinsically much fewer in number.

Finally, it is worth summarizing the primary assumptions we made in this work, where we have tried to err on the side of conservatism.  To simplify our calculations, we neglected emission from the unbound TDE debris, although that can substantially increase optical (non-super-Eddington) emission for low-mass SMBHs (SQ09).  Our SMBH wandering lifetimes were likely reduced by the fact that we limited ourselves to axisymmetric stellar bulge geometries, and even more importantly only considered spherical dark matter haloes.  Our simple choice of stellar mass function is slightly conservative for calculations of TDE rate.

\begin{table*}
 \begin{minipage}{144mm}
   \caption{Total number of offset TDEs observable by LSST per year ($\Gamma_{\rm s}$, $\Gamma_{\rm k}$) in different models.  The fiducial spin case is given first, and is bracketed in parentheses by the $a=0$ and $a=0.95$ cases.}
   \begin{tabular}{r|r|r|r|r|r|r|r|r|r}
     \hline
     Spin Alignment $[^{\circ}]$& super-Eddington? & $\Gamma_{\rm s} [\rm TDEs~yr^{-1}]$ & $\Gamma_{\rm k} [\rm TDEs~yr^{-1}]$ & $\Gamma_{\rm s, esc} [\rm TDEs~yr^{-1}]$ & $\Gamma_{\rm k, esc} [\rm TDEs~yr^{-1}]$ \\
     $10$& no & 0 (0, 0) & 0 (0, 0) & 0 (0, 0) & 0 (0, 0) \\
     $10$& yes & 0.7 (0.7, 0.7) & 0.5 (0.5, 0.5) & 0 (0, 0) & 0 (0, 0) \\
     $30$& no & 0.7 (0.3, 1.4) & 1.7 (0.2, 5.6) & 0 (0, 0) & 0 (0, 0) \\
     $30$& yes & 9.8 (9.5, 10)  & 5.9 (4.4, 9.7) & 1.0 (1.0, 1.0) & 0.8 (0.8, 0.8) \\
     $180$& no & 2.8 (0.4, 6.4) & 3.4 (0.3, 20) & 0.39 (0.3, 0.5) & 0.28 (0.2, 0.3) \\
     $180$& yes & 13 (11, 16) & 7.6 (4.5, 24) & 5.6 (5.6, 5.6) & 4.3 (4.3, 4.3) \\

   \end{tabular}
 \end{minipage}
\end{table*}

\section{Conclusions}
In this paper, we have demonstrated that super-Eddington flares from recoiling black holes, if they exist along the lines envisioned in SQ09, will be observably offset to LSST in numbers ranging from $\sim 1$ to $\sim 10$ TDEs per year.  This is true for a broad range of assumptions about kick velocity distributions and galactic structure.  This subset of transients would contain important scientific value as evidence of black hole recoil, and could potentially constrain the $v_{\rm k}$ distribution.  If super-Eddington flares do not exist or if they differ significantly from the SQ09 picture (for example, if $f_{\rm out} \ll 0.1$), optical emission from the accretion disks of TDEs around recoiling black holes will still be accessible to LSST, although here the case is more marginal.  If a large fraction of local universe SMBH mergers proceed without significant spin alignment, the prospects for optical detection of disk emission from recoiling TDEs are relatively good, but moderate amounts of alignment would likely suppress this.  Importantly, the majority of recoiled SMBHs will remain bound to their host galaxies, making photometric subtraction critical for identification of recoil-associated disruption flares.  Depending on the nature of the super-Eddington phase of accretion, a comparably large population of kinematically offset flares is potentially detectable, but would require spectroscopic followup to be realized.

We have also shown that confusion with TDEs from stationary SMBHs will not be a major challenge in the detection of off-nuclear TDEs, leaving supernova contamination as the main concern.  If the scientific potential of spatially offset TDEs is to be utilized, it will be necessary to construct transient survey pipelines which do not employ the typical ``galactic center'' cut when searching for TDEs.  Although the challenges inherent to TDE identification have been discussed elsewhere (van Velzen et al 2010), the distinctive lightcurve and color evolution of tidal disruption flares are helpful in separating them.  The large number of TDEs expected to be observed by time domain surveys in the coming decade will calibrate our understanding of these events, so that once LSST is online, it may be able to confirm the SMBH recoil predictions of numerical relativity.

\section*{Acknowledgments}
We thank Laura Blecha, Bence Kocsis and Ryan O'Leary for helpful suggestions.  This work was supported in part by NSF grant AST-0907890 and NASA grants NNX08AL43G and NNA09DB30A.


\begin{thebibliography}{99}

\bibitem[Baker et al.(2006)]{BCCK} Baker, J.~G., Centrella, 
J., Choi, D.~I., Koppitz, M., 
\& van Meter, J.\ 2006, Physical Review Letters, 96, 111102
\bibitem[Bandara et al.(2009)]{2009ApJ...704.1135B} Bandara, K., Crampton, 
D., \& Simard, L.\ 2009, ApJ, 704, 1135
\bibitem[Beloborodov et al.(1992)]{1992MNRAS.259..209B} Beloborodov, A.~M., 
Illarionov, A.~F., Ivanov, P.~B., 
\& Polnarev, A.~G.\ 1992, MNRAS, 259, 209 
\bibitem[Berti 
\& Volonteri(2008)]{2008ApJ...684..822B} Berti, E., \& Volonteri, M.\ 2008, ApJ, 684, 822
\bibitem[Binney 
\& Tremaine(2008)]{2008gady.book.....B} Binney, J., \& Tremaine, S.\ 2008, Galactic Dynamics: Second Edition, by James Binney and Scott Tremaine.~ISBN 978-0-691-13026-2 (HB).~Published by Princeton University Press, Princeton, NJ USA, 2008
\bibitem[Blecha et al.(2011)]{2011MNRAS.tmp...38B} Blecha, L., Cox, T.~J., 
Loeb, A., \& Hernquist, L.\ 2011, MNRAS, 38 
\bibitem[Blecha 
\& Loeb(2008)]{2008MNRAS.390.1311B} Blecha, L., \& Loeb, A.\ 2008, MNRAS, 390, 1311 
\bibitem[Bogdanovi{\'c} et al.(2004)]{2004ApJ...610..707B} Bogdanovi{\'c}, 
T., Eracleous, M., Mahadevan, S., Sigurdsson, S., 
\& Laguna, P.\ 2004, ApJ, 610, 707 
\bibitem[Bogdanovi{\'c} et al.(2007)]{2007ApJ...661L.147B} Bogdanovi{\'c}, 
T., Reynolds, C.~S., \& Miller, M.~C.\ 2007, ApJ, 661, L147
\bibitem[Bogdanovi{\'c} et al.(2009)]{2009ApJ...697..288B} Bogdanovi{\'c}, 
T., Eracleous, M., \& Sigurdsson, S.\ 2009, ApJ, 697, 288
\bibitem[Brenneman 
\& Reynolds(2006)]{2006ApJ...652.1028B} Brenneman, L.~W., \& Reynolds, C.~S.\ 2006, ApJ, 652, 1028 
\bibitem[Byun et al.(1996)]{1996AJ....111.1889B} Byun, Y.~I., et al.\ 1996, 
AJ, 111, 1889 
\bibitem[Campanelli et al.(2006)]{2006PhRvL..96k1101C} Campanelli, M., 
Lousto, C.~O., Marronetti, P., 
\& Zlochower, Y.\ 2006, Physical Review Letters, 96, 111101
\bibitem[Dehnen(1993)]{1993MNRAS.265..250D} Dehnen, W.\ 1993, MNRAS, 265, 
250 
\bibitem[Chandrasekhar(1943)]{1943ApJ....97..255C} Chandrasekhar, S.\ 1943, 
ApJ, 97, 255 
\bibitem[Chen et al.(2009)]{2009ApJ...697L.149C} Chen, X., Madau, P., 
Sesana, A., \& Liu, F.~K.\ 2009, ApJ Letters, 697, L149 
\bibitem[Chen et al.(2011)]{2011ApJ...729...13C} Chen, X., Sesana, A., 
Madau, P., \& Liu, F.~K.\ 2011, ApJ, 729, 13
\bibitem[Civano et al.(2010)]{2010ApJ...717..209C} Civano, F., et al.\ 
2010, ApJ, 717, 209 
\bibitem[Diener et al.(1995)]{1995MNRAS.275..498D} Diener, P., Kosovichev, 
A.~G., Kotok, E.~V., Novikov, I.~D., 
\& Pethick, C.~J.\ 1995, MNRAS, 275, 498
\bibitem[Dotti et al.(2009)]{2009arXiv0910.5729D} Dotti, M., Volonteri, M., 
Perego, A., Colpi, M., Ruszkowski, M., \& Haardt, F.\ 2009, arXiv:0910.5729
\bibitem[Escala et al.(2004)]{2004ApJ...607..765E} Escala, A., Larson, 
R.~B., Coppi, P.~S., \& Mardones, D.\ 2004, ApJ, 607, 765 
\bibitem[Evans 
\& Kochanek(1989)]{1989ApJ...346L..13E} Evans, C.~R., \& Kochanek, C.~S.\ 1989, ApJ, 346, L13
\bibitem[Faber et al.(1997)]{1997AJ....114.1771F} Faber, S.~M., et al.\ 
1997, AJ, 114, 1771 
\bibitem[Ferrarese 
\& Ford(2005)]{2005SSRv..116..523F} Ferrarese, L., \& Ford, H.\ 2005, Space Science Reviews, 116, 523
\bibitem[Gezari et al.(2008)]{2008ApJ...676..944G} Gezari, S., et al.\ 
2008, ApJ, 676, 944 
\bibitem[Gezari et al.(2009)]{2009astro2010S..88G} Gezari, S., et al.\ 
2009, astro2010: The Astronomy and 
Astrophysics Decadal Survey, 2010, 88 
\bibitem[Gonz{\'a}lez et al.(2007)]{2007PhRvL..98i1101G} Gonz{\'a}lez, 
J.~A., Sperhake, U., Br{\"u}gmann, B., Hannam, M., 
\& Husa, S.\ 2007, Physical Review Letters, 98, 091101 
\bibitem[Gualandris 
\& Merritt(2008)]{2008ApJ...678..780G} Gualandris, A., \& Merritt, D.\ 2008, ApJ, 678, 780
\bibitem[Guillochon et al.(2009)]{2009ApJ...705..844G} Guillochon, J., 
Ramirez-Ruiz, E., Rosswog, S., \& Kasen, D.\ 2009, ApJ, 705, 844 
\bibitem[Herrmann et al.(2007)]{2007PhRvD..76h4032H} Herrmann, F., Hinder, 
I., Shoemaker, D.~M., Laguna, P., \& Matzner, R.~A.\ 2007, PRD, 76, 084032 
\bibitem[Hills(1988)]{1988Natur.331..687H} Hills, J.~G.\ 1988, Nature, 331, 
687 
\bibitem[Hoffman 
\& Loeb(2007)]{2007MNRAS.377..957H} Hoffman, L., \& Loeb, A.\ 2007, MNRAS, 377, 957
\bibitem[Hopkins et al.(2007)]{2007ApJ...669...67H} Hopkins, P.~F., 
Hernquist, L., Cox, T.~J., Robertson, B., 
\& Krause, E.\ 2007, ApJ, 669, 67 
\bibitem[Hopkins et al.(2007b)]{2007ApJ...654..731H} Hopkins, P.~F., 
Richards, G.~T., \& Hernquist, L.\ 2007b, ApJ, 654, 731 
\bibitem[Hopkins et al.(2009)]{2009arXiv0906.5357H} Hopkins, P.~F., et al.\ 
2009, arXiv:0906.5357
\bibitem[Hunter 
\& Qian(1993)]{1993MNRAS.262..401H} Hunter, C., \& Qian, E.\ 1993, MNRAS, 262, 401 
\bibitem[Ivanov et al.(2005)]{2005MNRAS.358.1361I} Ivanov, P.~B., Polnarev, 
A.~G., \& Saha, P.\ 2005, MNRAS, 358, 1361 
\bibitem[Kasen 
\& Ramirez-Ruiz(2010)]{2010ApJ...714..155K} Kasen, D., \& Ramirez-Ruiz, E.\ 2010, ApJ, 714, 155 
\bibitem[Kesden et al.(2010)]{2010ApJ...715.1006K} Kesden, M., Sperhake, 
U., \& Berti, E.\ 2010, ApJ, 715, 1006 
\bibitem[King et al.(2005)]{2005MNRAS.363...49K} King, A.~R., Lubow, S.~H., 
Ogilvie, G.~I., \& Pringle, J.~E.\ 2005, MNRAS, 363, 49 
\bibitem[Komossa(2002)]{2002RvMA...15...27K} Komossa, S.\ 2002, Reviews in 
Modern Astronomy, 15, 27 
\bibitem[Komossa 
\& Merritt(2008)]{2008ApJ...683L..21K} Komossa, S., \& Merritt, D.\ 2008, ApJ, 683, L21
\bibitem[Komossa et al.(2008)]{2008ApJ...678L..81K} Komossa, S., Zhou, H., 
\& Lu, H.\ 2008, ApJ, 678, L81
\bibitem[Kormendy 
\& Bender(2009)]{2009ApJ...691L.142K} Kormendy, J., \& Bender, R.\ 2009, ApJ, 691, L142
\bibitem[Kormendy et al.(2009)]{2009ApJS..182..216K} Kormendy, J., Fisher, 
D.~B., Cornell, M.~E., \& Bender, R.\ 2009, ApJS, 182, 216
\bibitem[Kouwenhoven et 
al.(2007)]{2007A&A...474...77K} Kouwenhoven, M.~B.~N., Brown, A.~G.~A., Portegies Zwart, S.~F., \& Kaper, L.\ 2007, AAP, 474, 77 
\bibitem[Lauer et al.(1995)]{1995AJ....110.2622L} Lauer, T.~R., et al.\ 
1995, AJ, 110, 2622 
\bibitem[Lauer et al.(2005)]{2005AJ....129.2138L} Lauer, T.~R., et al.\ 
2005, AJ, 129, 2138 
\bibitem[Liu et al.(2009)]{2009ApJ...706L.133L} Liu, F.~K., Li, S., 
\& Chen, X.\ 2009, ApJL, 706, L133 
\bibitem[Lodato 
\& Pringle(2006)]{2006MNRAS.368.1196L} Lodato, G., \& Pringle, J.~E.\ 2006, MNRAS, 368, 1196 
\bibitem[Loeb(2007)]{2007PhRvL..99d1103L} Loeb, A.\ 2007, Physical Review 
Letters, 99, 041103
\bibitem[Loeb 
\& Ulmer(1997)]{1997ApJ...489..573L} Loeb, A., \& Ulmer, A.\ 1997, ApJ, 489, 573 
\bibitem[Lousto et al.(2010)]{2010CQGra..27k4006L} Lousto, C.~O., 
Campanelli, M., Zlochower, Y., 
\& Nakano, H.\ 2010, Classical and Quantum Gravity, 27, 114006 
\bibitem[Lousto et al.(2009)]{2009arXiv0910.3197L} Lousto, C.~O., Nakano, 
H., Zlochower, Y., \& Campanelli, M.\ 2009, arXiv:0910.3197 
\bibitem[LSST Science Collaborations et al.(2009)]{2009arXiv0912.0201L} 
LSST Science Collaborations, et al.\ 2009, arXiv:0912.0201 
\bibitem[Madau 
\& Quataert(2004)]{2004ApJ...606L..17M} Madau, P., \& Quataert, E.\ 2004, ApJ, 606, L17
\bibitem[Magorrian 
\& Tremaine(1999)]{1999MNRAS.309..447M} Magorrian, J., \& Tremaine, S.\ 1999, MNRAS, 309, 447
\bibitem[Merritt 
\& Milosavljevi{\'c}(2005)]{2005LRR.....8....8M} Merritt, D., \& Milosavljevi{\'c}, M.\ 2005, Living Reviews in Relativity, 8, 8 
\bibitem[Merritt et al.(2009)]{2009ApJ...699.1690M} Merritt, D., 
Schnittman, J.~D., \& Komossa, S.\ 2009, ApJ, 699, 1690 
\bibitem[Mihos 
\& Hernquist(1996)]{1996ApJ...464..641M} Mihos, J.~C., \& Hernquist, L.\ 1996, ApJ, 464, 641 
\bibitem[Navarro et al.(1997)]{1997ApJ...490..493N} Navarro, J.~F., Frenk, 
C.~S., \& White, S.~D.~M.\ 1997, ApJ, 490, 493 
\bibitem[O'Leary 
\& Loeb(2009)]{2009MNRAS.395..781O} O'Leary, R.~M., \& Loeb, A.\ 2009, MNRAS, 395, 781
\bibitem[O'Leary 
\& Loeb(2011)]{2011arXiv1102.3695O} O'Leary, R.~M., \& Loeb, A.\ 2011, arXiv:1102.3695 
\bibitem[{\"O}pik(1924)]{1924PTarO..25f...1O} {\"O}pik, E.\ 1924, 
Publications of the Tartu Astrofizica Observatory, 25, 6 
\bibitem[Ostriker(1999)]{1999ApJ...513..252O} Ostriker, E.~C.\ 1999, ApJ, 
513, 252 
\bibitem[Phinney(1989)]{1989IAUS..136..543P} Phinney, E.~S.\ 1989, The 
Center of the Galaxy, 136, 543 
\bibitem[NR(2007)]{NR2007} Press, W.~H. et al. 2007, Numerical Recipes. Cambridge University Press, Cambridge.
\bibitem[Pretorius(2005)]{Pre} Pretorius, F.\ 2005, 
Physical Review Letters, 95, 121101
\bibitem[Rees(1988)]{1988Natur.333..523R} Rees, M.~J.\ 1988, Nature, 333, 523 
\bibitem[Schnittman(2007)]{2007ApJ...667L.133S} Schnittman, J.~D.\ 2007, 
ApJL, 667, L133 
\bibitem[Schnittman 
\& Buonanno(2007)]{2007ApJ...662L..63S} Schnittman, J.~D., \& Buonanno, A.\ 2007, ApJ, 662, L63
\bibitem[Shields et al.(2009)]{2009arXiv0907.3470S} Shields, G.~A., et al.\ 
2009, arXiv:0907.3470 
\bibitem[Sijacki et al.(2010)]{2010arXiv1008.3313S} Sijacki, D., Springel, 
V., \& Haehnelt, M.\ 2010, arXiv:1008.3313 
\bibitem[Spergel et al.(2007)]{2007ApJS..170..377S} Spergel, D.~N., et al.\ 
2007, ApJS, 170, 377 
\bibitem[Stone 
\& Loeb(2011)]{2011MNRAS.412...75S} Stone, N., \& Loeb, A.\ 2011, MNRAS, 412, 75  
\bibitem[Strubbe 
\& Quataert(2009)]{2009arXiv0905.3735S} Strubbe, L.~E., \& Quataert, E.\ 2009, arXiv:0905.373
\bibitem[Syer 
\& Ulmer(1999)]{1999MNRAS.306...35S} Syer, D., \& Ulmer, A.\ 1999, MNRAS, 306, 35
\bibitem[Takahashi et al.(2010)]{2010SPIE.7732E..27T} Takahashi, T., et 
al.\ 2010, ProcSPIE, 7732  
\bibitem[Tremaine et al.(1994)]{1994AJ....107..634T} Tremaine, S., 
Richstone, D.~O., Byun, Y.-I., Dressler, A., Faber, S.~M., Grillmair, C., 
Kormendy, J., \& Lauer, T.~R.\ 1994, AJ, 107, 634  
\bibitem[Tremaine et al.(2002)]{2002ApJ...574..740T} Tremaine, S., et al.\ 
2002, ApJ, 574, 740 
\bibitem[Toomre(1964)]{1964ApJ...139.1217T} Toomre, A.\ 1964, ApJ, 139, 
1217 
\bibitem[van Velzen et al.(2010)]{2010arXiv1009.1627V} van Velzen, S., 
Farrar, G.~R., Gezari, S., Morrell, N., Zaritsky, D., Ostman, L., Smith, 
M., \& Gelfand, J.\ 2010, arXiv:1009.1627 
\bibitem[Vereshchagin et 
al.(1988)]{1988Ap&SS.142..245V} Vereshchagin, S., Tutukov, A., Iungelson, L., Kraicheva, Z., \& Popova, E.\ 1988, APSS, 142, 245 
\bibitem[Volonteri et al.(2005)]{2005ApJ...620...69V} Volonteri, M., Madau, 
P., Quataert, E., \& Rees, M.~J.\ 2005, ApJ, 620, 69 
\bibitem[Wang 
\& Merritt(2004)]{2004ApJ...600..149W} Wang, J., \& Merritt, D.\ 2004, ApJ, 600, 149
\bibitem[Wegg 
\& Bode(2010)]{2010arXiv1011.5874W} Wegg, C., \& Bode, N.\ 2010, arXiv:1011.5874
\bibitem[Wetzel 
\& White(2009)]{2009arXiv0907.0702W} Wetzel, A.~R., \& White, M.\ 2009, arXiv:0907.0702 
\bibitem[Yu 
\& Tremaine(2003)]{2003ApJ...599.1129Y} Yu, Q., \& Tremaine, S.\ 2003, ApJ, 599, 1129 

\end{thebibliography}
\end{document}